\documentclass[prd,twocolumn,nofootinbib,amsmath]{revtex4}

\usepackage{amsmath,amssymb}
\usepackage{graphicx,units}

\newcommand\NOTE[1]{\mbox{}\\\textbf{--- #1 ---}\\}
\def\??#1{\NOTE{#1}}

\def\xleft{\mathopen{}\left}

\DeclareMathOperator{\cov}{cov}

\begin{document}

\title{Two-population model for MTL neurons: The vast majority are
  almost silent}

\author{Andrew Magyar}
\author{John Collins}
\affiliation{Physics Department,
   Pennsylvania State University,
   University Park, PA 16802, USA
}
\date{07 November 2014}

\begin{abstract}
  Recordings in the human medial temporal lobe have found many neurons
  that respond to pictures (and related stimuli) of just one
  particular person out of those presented.  It has been proposed that
  these are concept cells, responding to just a single concept.
  However, a direct experimental test of the concept
  cell idea appears impossible, because it would need the measurement of the
  response of each cell to enormous numbers of other stimuli.  Here we
  propose a new statistical method for analysis of the data, 
  that gives a more powerful way to analyze how close data are to the
  concept-cell idea.
  It
  exploits the large number of sampled neurons, to give sensitivity to
  situations where the average response sparsity is to much less than
  one response for the number of presented stimuli.  We show that a
  conventional model where a single sparsity is postulated for all
  neurons gives an extremely poor fit to the data. In contrast a model
  with two dramatically different populations give an excellent fit to
  data from the hippocampus and entorhinal cortex. In the hippocampus,
  one population has 7\% of the cells with a 2.6\% sparsity. But a
  much larger fraction 93\% respond to only 0.1\% of the stimuli. This
  results in an extreme bias in the reported responsive of neurons
  compared with a typical neuron.  Finally, we show how to allow for
  the fact that some of reported identified units correspond to
  multiple neurons, and find that our conclusions at the neural level
  are quantitatively changed but strengthened, with an even stronger
  difference between the two populations.
\end{abstract}
  
\maketitle

\section{Introduction}
\label{sec:intro}

A long-standing debate (e.g., \cite{Bowers09, Bowers10a, Bowers10b,
  Plaut_McClelland_2010a, Plaut_McClelland_2010b,
  Quiroga_Kreiman_2010a, Quiroga_Kreiman_2010b}) concerns whether
biological neural 
systems use local or distributed coding\footnote{For our purposes, we
  will mean by a local code that there exist single neurons for
  particular concepts, ideas, or other entities, so that activity of
  one of these neurons above some threshold indicates exclusively the
  presence of the corresponding entity.  With distributed coding,
  measurements from several or many neurons are needed to determine
  whether a particular entity is present. }  for the representation of
high-level entities, like individual people.  Although the consensus
\cite{sparse.coding} is that only
distributed coding is used, a number of 
experiments find results that are very suggestive of local coding. For
example, Hahnloser, Kozhevnikov and Fee \cite{HVC-RA} find neurons in
zebra finches that each fire only at one particular time during a
bird's singing of its own song.  In a sensory context in the medial
temporal lobe (MTL) of human epileptic patients, Quian Quiroga et al.\
\cite{GMC} report many examples of neurons that fire only in response
to a stimulus that contains a particular person.  It has been
suggested that these neurons are concept cells \cite{concept.cells},
responding to one concept out of many possible. 

It is therefore important to measure how close such neurons are to the
local coding situation. There are some interesting conceptual issues
associated with what precisely should be meant by the previous
sentence, and we will discuss these further in Sec.\
\ref{sec:represention}.  Among these is the question of whether a
particular neuron should be considered as actually participating in
the representation of a particular concept, when it is measured that
the neuron responds strongly to the presence (in some sense) of that
concept in the stimulus.  It might be, for example, that, instead, the
neuron represents an episodic memory whose content includes the
concept.

For our immediate purposes, we do not actually have to solve this
issue.  The problem we address is that any neuron that does local
coding for a high-level concept will respond to only a very small
fraction of stimuli, so that direct experimental measurements have
great difficulty testing whether the coding is actually local.
Consider the classic case where the concepts represented are actually
those detected in a current stimulus.  For example, the concept might
be that of a particular individual human in a visual scene.  To
measure whether a particular cell implements local coding for a
concept, one would need presentation of stimuli that cover a large
fraction of the concepts known to the subject.  This is evidently far
beyond current experimental capabilities, at least.  Measurements such
as those in \cite{GMC, Mormann} use stimuli corresponding to only
about 100 distinct entities (people, famous buildings, etc).

As one of us has shown \cite{CJ} in collaboration with Jin, a
substantial complication in the interpretation of the data is that
there are wide differences in the response properties of the neurons
in question.  A small percentage respond to several distinct stimuli
out of those presented, while a vast majority respond on average to
much less than one.  This was quantitatively deduced from summary
statistics for the data that were given in \cite{GMC}.  (No further
breakdown of the data was given.)  We made a prediction for the
fraction of neurons as a function of the number of stimuli to which
they make an above-threshold response.

In this paper, we present and use a more complete version of the novel
statistical method underlying our earlier work, to deduce the
conceptual-coding properties of neurons.  Applying it to more recent
and more detailed data from \cite{Mormann}, we will find that in fact
the vast majority of the measured neurons respond on average to much
less than one stimulus in the approximately 100 that are presented.  A
direct measurement on a single neuron would need thousands of
conceptually distinct stimuli to achieve the same result.  The
effectiveness of our method arises from the large number of neurons
probed.  A measure of the method's power is the number of
neuron-stimulus trials, i.e., the number of stimuli times the number
of neurons.  This is in the hundreds of thousands for the data of
\cite{GMC, Mormann}.

From properties of the set of neuron-stimulus trials, treated as a
sample, we deduce estimated response properties for the ``universe''
of all neurons in particular brain regions and all relevant stimuli.
We will formulate the results in terms of the sparsity of individual
neurons. Our definition is that the sparsity of a particular neuron is
the fraction of stimuli to which it gives a response above some
appropriate threshold (e.g., as chosen in \cite{GMC, Mormann}).  Our
aim is to estimate the distribution of sparsity over neurons in a
particular brain region.  The simplest model, \cite{Waydo}, is that
all neurons have (approximately) the same sparsity.  But, as we will
show, such a model is in dramatic disagreement with the data; a more
general distribution with widely varying sparsities for different
neurons is needed, and we will estimate such a distribution, and show
that most neurons in the areas concerned have a sparsity below
$10^{-3}$.

These results confirm and extend those we made earlier \cite{CJ}.  In
particular, the prediction mentioned above is successfully tested,
with a quantitative measure of its (excellent) goodness of fit to the
newer data. (Minor changes in parameters are needed for the different
set of data.)

It has previously been pointed out, notably in \cite{dark.matter,
  Olshausen.Field}, that many neurons are unresponsive or even silent,
within the limits of experimental measurements over a limited period,
as is quantitatively supported by our results.  Hence a problem is
that reported measurements of responsive neurons can give a very
biased picture of the nature of neural coding.  Our method provides a
tool for both measuring the bias and for compensating for it. 

In our discussion section, Sec.\ \ref{sec:discuss}, we include remarks
on the implications of our results for the nature of neural coding of
high-level concepts.

\section{Outline}
\label{sec:outline}

After setting up our methods, we apply them to published data from
Mormann et al.\ \cite{Mormann}, which presents measurements of the
number of neurons as a function of the number of stimuli (out of a
total of around 100) to which a particular neuron gives an
above-threshold response.  The data are for four different areas of
the MTL pooled across several patients.

For the distribution of sparsity, we use a model whose general form
has multiple populations of cells, each characterized by a particular
sparsity and a fraction of the total neural population.  We regard an
experiment as using a random sample of neurons and stimuli.  The model
is in fact a particular kind of mixture model, in statistical
terminology, with one component of the mixture for each population. To
fit the model to data, an appropriate method is the maximum likelihood
method, and it allows us not only to measure the parameters of the
model and their uncertainties, but also to evaluate the goodness of
fit with the aid of a $\chi^2$ function.  We show that the plots given
in \cite{Mormann} (numbers of neurons responding a particular number
of stimuli) give sufficient statistics for fitting the model, in the
sense of standard statistical theory.  The much more elaborate
statistical treatment used by Waydo et al.\ \cite{Waydo} is actually
unnecessary, and, importantly, did not provide a measure of goodness
of fit.

One motivation for using populations of neurons each with a particular
value of sparsity arises from work by Attwell and Laughlin
\cite{metabolism2} and by Lennie \cite{metabolism}.  There it is shown
that to optimize the energy consumption by neurons transmitting a
given amount of information, a particular value of sparsity is
preferred, around a few percent.  Even if other constraints besides
energy consumption are important, the analyses suggest that sparsity
is a parameter that could be adjusted (e.g., by evolution) to optimize
the performance of a neural system.  Therefore it is sensible to
propose models in which particular populations of neurons have
particular values of sparsity.

Another motivation for trying a multiple-population model is that
recent experiments have detected multiple populations of neurons
located in the various song-related regions of the brains of zebra
finches \cite{HVC-RAX}.  These sets of neurons have strikingly
different properties of these sets of neurons as regards how often
they spike. 

We first show that the simplest possible model with one population of
neurons, all with the same sparsity, provides an unacceptably bad fit
to the data.  Such a model is quite often used, for example
\cite{Waydo} by the experimental group whose data we fit.

The nature of the deviations between the single-population model and
the data will show that at least one more population of very sparsely
responding neurons is needed.  As an intermediate step we show that a
better fit is obtained by adding a population of silent neurons, i.e.,
of neurons that gave no above threshold responses whatever in the
experiment.  This provides a greatly improved fit, as expected, but
the fit is still poor.  Interestingly, the silent neurons are in the
majority.

Our main model has two populations of neurons each with a particular
sparsity.  It has three parameters: the sparsities of the two
populations and the relative size of the two populations.  We will
find greatly improved fits.  In the case of the hippocampus, the fit
is completely acceptable by the appropriate $\chi^2$ criterion,
although in the other areas the fits are poorer, particularly for the
PHC.  In all cases, we find that at most only about 5\% of the neurons
have a normal sparsity of a few percent.  The remaining 95\% or more
of the neurons respond ultra-sparsely, with a sparsity around 0.1\%.
Thus the vast majority of the neurons are almost silent: they respond
on average to much less than one out of the approximately 100 distinct
stimuli presented in the experiments.  The neurons that are reported
as responding are therefore an extremely biased sample of all neurons
in the tested regions.  Our statistical methods provide a way to
quantify this bias and to compensate for it.

Our finding of a large number of silent cells quantitatively supports
the conclusions of Shoham, O'Connor and Segev \cite{dark.matter} about
neural dark matter.  It is similar, but more extreme, than a similar
conclusion by Olshausen and Field \cite{Olshausen.Field} about the
(apparent) silence of many neurons in area V1 of the visual cortex.

There are other possible definitions of sparsity and/or selectivity
besides the one we chose, and other definitions can give what appear
to be dramatically different conclusions.  A notable example is the
definition of Treves and Rolls \cite{Treves.Rolls91}, who
\cite{Rolls.Treves.2011} find much larger values of sparsity, around
30\%--40\%.  We will argue that, although the two definitions are
equivalent in a certain limit, the high values for Rolls-Treves
sparsity can be completely misleading as to the nature of neural
coding.  We will point out that it is possible that neurons could have
a relatively low number of spikes in response to many stimuli but a
dramatically higher and readily identifiable response to just a few
stimuli.  In that case, it can be that the Rolls-Treves sparsity
measures mostly the variability of the common low responses, while our
sparsity measures the fraction of the selective high responses.

Finally, we will comment on whether our conclusions are robust, on its
implications, and on the possibility raised by Waydo et al.\
\cite{Waydo} that there may be even more apparently silent neurons, so
that the true conclusions about ultra-sparsely responding neurons are
even more extreme by a substantial factor than what we have fitted.

\section{General multi-population model and its statistical analysis}
\label{sec:general.model}

We define the sparsity of a unit or neuron as the fraction of total
stimuli that elicit a response from that unit.  For our purpose
distinct stimuli are distinct visual images of people, objects, etc,
such as used in \cite{GMC}.  The threshold for defining a response is
set by the experimenters \cite{GMC}.  Only a very small subset of the
total possible stimuli are presented in a particular experimental
context, and any precise estimates of the sparsity distribution are
likely to depend on the circumstances of the experiment.

This definition of sparsity is appropriate where the measured neurons
typically have a low firing rate, and occasionally have a much higher
firing rate, under specific circumstances.
This applies to typical neurons in the kind of data we analyze ---
see examples in \cite{GMC,Mormann}.

Related examples that suggest when our definition is appropriate can
be found in neural activity in the HVC area of zebra finches during
their song \cite{HVC-RA,HVC-RAX}.  RA-projecting and X-projecting
neurons have occasional high firing rates at consistent points in a
bird's song, with few spikes elsewhere.  Our thresholded definition of
sparsity would be appropriate (with different values for the RA- and
X-projecting neurons, of course).  The high responses can be usefully
read out, not just by experimentalists, but by other ``reader''
neurons \cite{Buzsaki.2010}.  In contrast interneurons fire at a high
rate during much of the song, but with fairly consistent patterns of
various levels of activity.  For these, the thresholded, binary
definition of sparsity would be less appropriate.  It might well be
that the interneurons have very low sparsity, since deviations of
several standard deviations above the mean firing rate are likely to
be rare, according to a visual examination of the figures in
\cite{HVC-RA,HVC-RAX}.  But this would be misleading as to the nature
of the interneuron's firing properties.

In the data that we analyze from \cite{Mormann}, 
units were identified from electrode recordings using spike
sorting techniques which cannot always distinguish individual
neurons. Thus, the firing patterns reported for some units represent
the aggregate firing of multiple neurons rather than for a single
neuron.  The simplest version of our model ignores this distinction,
and assumes that each recorded unit consists of only a single
neuron. But, as we will explain in Sec.\ \ref{sec:multi}, the general
principles of our methods apply perfectly well at the unit level.  We
will show how to transform from a neural-level model to a unit-level
model, and we will see how to interpret numerical results of fits at
the unit level in terms of single neuron properties.  This will result
in no change in our qualitative results, but a strengthening of our
conclusions about the presence of a large number of very sparsely
responding neurons.


\subsection{Model Definition}
\label{sec:model.defn}

In the data we analyze, the unit firing patterns are treated in a
binary fashion. The threshold for a response is defined by the
experimentalists \cite{Mormann, GMC}. We assume: 
\begin{itemize}
	\item The sparsity of each neuron remains constant over the
          course of the experiment. 
	\item The recorded neurons are statistically independent of
          each other. 
	\item The neurons are partitioned into distinct populations,
          each with fractional abundance, $f_i$,
          where $i$ labels the population.
	\item All neurons in population $i$ have the same sparsity,
          $\alpha_i$. 
\end{itemize}
The statistical independence of the recorded neurons was verified
experimentally \cite{GMC}.

Our ultimate goal is find an appropriate set of populations, with
their fractional abundances and sparsities.  However, as will become
apparent later, data has limited ability to distinguish populations
with sparsities that are close to each other: in that situation the
effect is close to that of a single population with a single averaged
sparsity.  In contrast, the data do provide the ability to distinguish
populations of widely different sparsities.  So our practical goal
will be to find the simplest model that is consistent with reported
data; the model is thus meant only to be a useful approximation to
reality.

In this paper, we will start from a very simple model with one
population of neurons, and elaborate it in two stages.  This will give
three models:
\begin{enumerate}
\item Single-population model: All neurons have the same sparsity, $\alpha$. This model has one parameter.
\item Two-population model, with one population silent: The active
  population has sparsity $\alpha_{\rm D}$ and abundance $f_{\rm D}$, while
  the other population is silent with $\alpha_S=0$, and abundance $1-f_{\rm
    D}$. This model has two parameters.
\item Full two-population model: Two active populations are present,
  one with parameters $\alpha_{\rm D}$ and $f_{\rm D}$, and one
  ultra-sparse population with parameters $\alpha_{\rm US}$ and
  $f_{\rm US}=1-f_{\rm D}$. This model has three parameters. The labeling
  of the populations is defined by choosing $\alpha_{\rm US} < \alpha_{\rm D}$.
  
\end{enumerate}
For each model, we first produce the best fit to the data in
\cite{Mormann}, by using a maximum likelihood estimates (MLE) for the
parameters \cite{Lyons}.  We will do this for each of the four regions
of the MTL for which data is reported.  Then we test the goodness of
the fit by using a $\chi^2$ analysis.

\subsection{Mathematical characterization of data and model}

To implement the MLE of the parameters, we first derive the likelihood
function for the model.  This is defined \cite{Lyons} as the
probability of the data given the model and its parameters.

Given that the neurons are treated as binary (responsive or not
responsive), the data from an experiment in which $N$ neurons are
recorded during the presentation of $S$ stimuli can be fully
represented by a collection of $NS$ Bernoulli random variables,
$X_{js}$
\begin{equation}\label{eq.rawdata}
X_{js}=
\begin{cases}
   1, &  \text{if neuron $j$ responds to stimulus $s$}, \\
   0, &  \text{if neuron $j$ does not respond to $s$}.
\end{cases}
\end{equation}

Our model in its general form has $M$ populations, with population $i$
having fractional abundance $f_i$ and sparsity $\alpha_i$.  The fractional
abundances add to unity:
\begin{equation}
\label{normal}
\sum_{i=1}^{M}{f_i}=1,
\end{equation}
so that there are $2M-1$ model parameters: the sparsity of each
population, the fractional abundance of each population, but minus one
for the constraint (\ref{normal}).

If we knew the population $i_j$ to which a neuron $j$ belongs, then
the probability distribution of $X_{js}$ would be\footnote{Here and
  elsewhere, we use a notational convention that is common in the
  statistical literature: $X_{js}$ with an upper case $X$ denotes a
  random variable in the technical statistical sense, while the
  corresponding symbol $x_{js}$ with a lower-case name $x$ denotes a
  numerical value of the random variable that results from a
  particular set of experimental observations.}
\begin{align}\label{rawdist}
   \text{Prob}\xleft( X_{js}{=}x_{js} \mid \alpha_{i_j} \right)
  &=
   \begin{cases}
      \alpha_{i_j}, &  \text{if $x_{js}=1$}, \\
      1-\alpha_{i_j}, &  \text{if $x_{js}=0$}
   \end{cases}
\nonumber\\
   &= 
   \alpha_{i_j}^{x_{js}}\left(1-\alpha_{i_j}\right)^{1-x_{js}}.
\end{align}
Then the probability of a particular outcome in the showing of $S$
stimuli to the whole set of neurons is:
\begin{align}
\label{data}
\text{Prob} \xleft( \{X_{js}=x_{js}\} \mid \{\alpha_{i_j}\} \right)
& =
\prod_{j,s} \text{Prob}\xleft( X_{js}{=}x_{js} \mid \alpha_{i_j} \right)
\nonumber\\
& \hspace*{-20mm}
 = \prod_j \xleft[
 \alpha_{i_j}^{\sum_{s=1}^{S}{{x_{js}}}}
 \left(1-\alpha_{i_j}\right)^{S-\sum_{s=1}^{S}{{x_{js}}}}
 \right].
\end{align}
Here $\{X_{js}=x_{js}\}$ and $\{\alpha_{i_j}\}$ denote the whole array of
the quantities notated.

It is now convenient to define two sets of auxiliary random
variables.  One is the number of responses that a particular neuron
makes:
\begin{equation}
  \label{eq:Kj.defn}
  K_j = \sum_{\text{stimuli }s} X_{js}.
\end{equation}
The second is the number of neurons $N_k$ that give $k$ responses to
the stimuli:
\begin{equation}
  \label{eq:Nk.def}
  N_k = \sum_{\text{neurons }j} \delta_{k,K_j},
\end{equation}
where as usual, the Kronecker delta $\delta_{\alpha\beta}$
obeys $\delta_{\alpha\beta}=1$ if $\alpha=\beta$ and
$\delta_{\alpha\beta}=0$ otherwise. 

We can then write the probability of the data (given the
set of $\alpha_{i_j}$) in terms of the values $k_j$ of the random
variables $K_j$ alone: 
\begin{equation}
\label{data2}
\text{Prob} \xleft( \{X_{js}=x_{js}\} \mid \{\alpha_{i_j}\} \right)
 = \prod_j \xleft[
     \alpha_{i_j}^{k_j}
     \left(1-\alpha_{i_j}\right)^{S-k_j}
 \right].
\end{equation}
But we do not know the values of each neuron's sparsity, so the
probability of the data given the model parameters is given by summing
over the possible sparsity values weighted by their probabilities:
\begin{multline}
\label{data3}
\text{Prob} \xleft( \{X_{js}=x_{js}\} \mid \{\alpha_i, f_i\} \right)
\\
= \prod_{j=1}^N
   \xleft[
     \sum_{i=1}^M f_i
     \alpha_i^{k_j}
     \left(1-\alpha_i\right)^{S-k_j}
 \right].
\end{multline}
This depends only on the values of the random variables $K_j$, and not
on any more 
detailed properties of the data.  That is, the values of $K_j$ are
sufficient statistics for fitting the model.

Therefore it is useful to compute the probabilities for the $K_j$.
First, for one neuron of given sparsity $\alpha$, the probability of $k$
responses, in the presentation of $S$ random images, is the binomial
distribution:
\begin{equation}
\label{single neuron}
P(K=k \mid \alpha)
= \binom{S}{k} \, \alpha ^{k} (1-\alpha) ^{S-k}.
\end{equation}
Here the factor $\binom{S}{k} = S!/(k!\, (S-k)!)$ counts the number of
ways of getting $k$ responses to $S$ stimuli.  The distribution
(\ref{single neuron}) has mean $\alpha S$ and standard deviation
$\sqrt{\alpha(1-\alpha)S}$, which is approximated by $\sqrt{\alpha S}$
for the actual situation of $\alpha\ll1$.  The ratio of standard
deviation to mean is $\sqrt{(1-\alpha)/(\alpha S)} \simeq
1/\sqrt{\alpha S}$, which goes to zero as $S\to\infty$.  If we plot
the distribution of $k/S$, i.e., the distribution of the fractional
response rate, this can be regarded as a smearing of the delta
function $\delta(k/S-\alpha)$.

In the model, there are $M$ distinct neuronal populations, each with
sparsity $\alpha_i$ and fractional abundance $f_i$.  If the
population to which the neuron belongs is unknown, then its
probability of responding to $k$ out of $S$ stimuli is
\begin{align}
\label{random.neuron.multi}
 \epsilon_{k} \equiv
 P(K{=}k) & = \sum_i { f_i P(K{=}k \mid \alpha_i) } \
\nonumber\\
  &= \sum_{i=1}^{M}{f_i \binom{S}{k} \alpha ^{k}_i
  (1-\alpha_i) ^{S-k}}.
\end{align}
With the aid of Eq.\ (\ref{normal}), one can check that the
probabilities in Eq.\ (\ref{random.neuron.multi}) correctly sum
to unity: $\sum_{k=0}^S \epsilon_k = 1$.

We already know that the $K_j$ are sufficient statistics for fitting
the model to the data.  So we need the probabilities for the set of
$K_j$:
\begin{equation}
\label{eq:P.K}
\text{Prob}\xleft(\bigwedge_{j=1}^{N} K_j=k_j \right)
  =
    \prod_{j=1}^{N}{\epsilon_{k_j}}
  = \prod_{k=0}^{S}{\epsilon_{k}^{n_k}},
\end{equation}
where ``$\land$'' denotes ``and''.  In the last part of this equation, we
have simply counted the number of neurons with a given value of $k$,
for every possible value of $k$.  Since the result depends only on the
values of $n_k$, these themselves form a set of sufficient statistics
for fitting the model to the data. Ref.\ \cite{Mormann} provided
results for these numbers, and no further information on the raw
experimental data is needed to fit the parameters of the model;
examination of the other quantities used in the work of Waydo et al.\
\cite{Waydo} is unnecessary.

\subsection{Likelihood and its analysis}

Since the random variables $N_k$ are sufficient statistics, we need
the probabilities for them.  From these we obtain the likelihood
function that we will use for fitting the model to the data. It is
obtained from Eq.\ (\ref{eq:P.K}) by counting the number of different
arrays of $K_j$ that given a given set of $N_k$s.  We then get
\begin{multline}
\label{likelihood}
    \mathcal{L}\xleft( \{ \alpha_i, f_i \} \mid \text{Data} \right)
  \equiv
    \text{Prob}\xleft( \{N_k=n_k\}\vert \{\alpha_i, f_i\} \right)
\\
  =
    \frac{N!}{\prod_{k=0}^{S}{n_k!}}\prod_{k=0}^{S}{\epsilon_{k}^{n_k}}.
\end{multline}
In this likelihood function, the dependence on the model parameters,
$\{ \alpha_i, f_i \}$, is contained in the $\epsilon_k$.

Note that in making the transformation from a distribution over
$X_{js}$ to $K_j$ and then to $N_k$, we have greatly reduced the
number of data items to be considered, from $NS$ to $N$ (the number of
neurons) to $S$ (the number of stimuli).  Moreover, as can be seen
from the data in \cite{Mormann}, only the first few $N_k$ are nonzero
in reality, at least to a good approximation.

To estimate the values of the parameters of the model, we use the
maximum-likelihood method, as is appropriate for this situation. 
 
The methods we use are, in fact, almost identical to long-established
methods \cite{Baker.Cousins, Lyons} that are regularly used for
analyzing data from scattering experiments in high energy physics (and
more generally, scattering experiments in physics).  This close
similarity arises because both in the scattering experiments and in
the neural data, we have a large number of independent trials, and the
probability of a non-trivial outcome is small.  A non-trivial outcome
in the physics case is a scattering event between pairs of particles
in the beams in an experiment, while in the neural case it is an
above-threshold response by a particular neuron to a particular
stimulus.  The physics analog of the neural $N_k$ is the number of
scattering events in a certain bin of kinematics.

Although our methods are informed by those used in the physics case,
we will provide a self-contained treatment appropriate to the neural
case.  An important tool we will take over from high-energy physics is
an appropriately defined $\chi^2$ function that is closely related to
the logarithm of the 
likelihood function.  Minimization of a suitably defined
\cite{Baker.Cousins, Hauschild.Jentzel} $\chi^2$ is equivalent to
maximizing likelihood.  The minimum value of $\chi^2$ provides a very
convenient measure of goodness of fit, to assess how consistent the
data are with the model.  The shape of the $\chi^2$ function near the
minimum can also be used to estimate the uncertainties on the fitted
values of the parameters of the model, and also the correlations in
the uncertainties.

The $\epsilon_{k}$ are subject to the normalization constraint
\begin{equation}\label{epnormal}
\sum_{k=0}^{S}{\epsilon_k}=1,
\end{equation}
while the $n_k$ are constrained by
\begin{equation}\label{Nsum}
\sum_{k=0}^{S}{n_k}=N.
\end{equation}
The maximum likelihood method finds the best fit to the data by
maximizing $\mathcal{L}\xleft( \{ \alpha_i, f_i \} \right)$ with
respect to the parameters $f_i$ and $\alpha_i$.

The data show that the probability of a neural response is much less
than one.  Thus we can usefully make approximations appropriate for
the situation that $\epsilon_k \ll 1$ and $n_k \ll N$ for $k\geq 1$.
Hence $\epsilon_0$ is close to unity, while $n_0$ is less than $N$
only by a small fraction.
Under these approximations, the likelihood function
simplifies to:
\begin{align}
\label{likelihoodfinal}
   \mathcal{L}
   \approx
   \prod_{k=1}^{S} 
      e^{-N \epsilon_{k}} \frac{\left(N \epsilon_{k}\right)^{n_k}}{n_k!},
\end{align}%
i.e., a product of Poisson distributions for each of the
\emph{non-zero} $k$ values.  A derivation is given in the
Appendix. 

\subsection{Error analysis}
\label{sec:error.analysis}

Suppose that we have estimated the best fit parameters by maximizing
likelihood with respect to the parameters $\{\alpha_i,f_i\}$.  Then
taking the likelihood function to be approximately Gaussian in the
vicinity of its maximum, we can estimate the confidence intervals and
correlations of the model parameters \cite{Lyons} from the diagonal
terms of the covariance matrix, defined by:
\begin{equation}\label{covariance}
\cov\left(\theta_i,\theta_j\right)=-\left(H^{-1}\right)_{ij},
\end{equation}%
where the Hessian matrix is
\begin{equation}\label{hessian}
H_{ij}=\frac{\partial^2}{\partial\theta_i \partial\theta_j}\ln\mathcal{L}.
\end{equation}%
The diagonal terms yield the parameter variances, $\cov\left(\theta_i,\theta_i\right)=\sigma_i^2$. Correlations between model parameters, $\rho_{ij}$, can be determined from the off-diagonal elements of Eq.\ (\ref{covariance}),%
\begin{equation}\label{correlations}
\rho_{ij}=\frac{\cov\left(\theta_i,\theta_j\right)}{\sigma_i\sigma_j}.
\end{equation}%

Two important properties of a Poisson distribution $e^{-N
  \epsilon_{k}} \left(N \epsilon_{k}\right)^{n_k} / n_k! $ are its
mean and standard deviation
\begin{equation}
  \langle N_k \rangle = N\epsilon_k,
\qquad 
  \text{s.d.}( N_k ) = \sqrt{N\epsilon_k}.
\end{equation}
Thus we can characterize the typical values of $n_k$ by $n_k =
N\epsilon_k \pm \sqrt{N\epsilon_k}$. 

We will assess goodness of fit by using the following $\chi^2$ function:
\begin{equation}
  \label{eq:chi2}
  \chi^2(k_{\rm max}; n_1, n_2, \dots)
  = \sum_{k=1}^{k_{\rm max}} \frac{ (n_k-N\epsilon_k)^2 }{ N\epsilon_k }.
\end{equation}
This corresponds to $-2 \ln \mathcal{L}$, to within an additive
constant, when the Poisson distributions are replaced by Gaussian
approximations near their peak.  
But this approximation is only suitable when
$N\epsilon_k$ is reasonably much larger than one.  So, in Eq.\
(\ref{eq:chi2}), we truncated the sum over $k$ to $k\leq k_{\rm max}$.
The restriction should be those values of $k$ such that the expected
value of $n_k$ is
bigger than one or two, i.e., where there are noticeable neural
responses.  Beyond $k_{\rm max}$, there are very few neural responses,
and therefore little information for fitting the parameters of the
model. 

After we have we performed a maximum-likelihood estimate of the
parameters of the model, statistical theory \cite{Lyons} predicts a
mean and standard deviation for $\chi^2$:
\begin{equation}
  \label{eq:chi2.value}
  \chi^2 = N_{\rm dof} \pm \sqrt{ 2 N_{\rm dof} }.
\end{equation}
where the number of degrees of freedom, $N_{\rm dof}$, is the number
of data values used (i.e., the number of values of $k$ in the
truncated sum) minus the number of parameters fitted (which will be 1,
2 or 3, in the particular implementations of the model that we
use). If the value of $\chi^2$ falls much outside this range, that
indicates that the model does not agree with the data.

\section{Multi-Unit Considerations}
\label{sec:multi.basics}

The preceding analysis assumes that the recorded units all consist of
a single neuron. However, only a fraction of the recorded units
contain a single neuron while the rest are composed of several
neurons. 

We now show that given our general multipopulation model at the neuron
level, a version of the model also applies at the unit level.  That
is, each unit has a sparsity, which can be calculated as a function of
the sparsities of its constituent neurons, and there are populations
of units with different values of sparsity.  

This is the general picture. Some simplifications and useful
approximations can be made in the application to real data, as we will
see in Sec.\ \ref{sec:multi}.

Suppose first that a particular unit consists of $R$ neurons of known
sparsities.  Let $r=1,\dots,R$ label the neurons, and let $\alpha_{i_r}$
be the sparsity of neuron $r$. The unit's sparsity,
i.e., the probability that the unit responds to a presented stimulus
is given by:
\begin{equation}
\label{eq.multiunit.sparsity}
   \alpha' \equiv
   \text{Prob}\left(\text{response} \mid R,
                    \alpha_{i_1},\dots,\alpha_{i_R}
              \right)
   = 1- \prod_{r=1}^{R}{\left(1-\alpha_{i_r}\right)}.
\end{equation}
Then the probability that the unit responds to $k$ of $S$ stimuli given $R$ and the $\alpha_{i_r}$ is a binomial distribution:
\begin{equation} 
\label{eq.multiunit.binomial}
	\text{Prob}(K=k \hspace{1 mm} | \hspace{1 mm} R,\alpha_{i_1},\dots,\alpha_{i_R})=\binom{S}{k} \hspace{0.5 mm} (\alpha' )^{k} \hspace{0.5 mm} (1-\alpha') ^{S-k}.
\end{equation} 

So far, we have assumed that $R$ and the sparsities $\alpha_{i_R}$ are
known.  Next, we assess the unit sparseness in terms of the
probability distribution of the neural sparsities.
Let $\epsilon^{\rm unit}_{k,R}$ be the total probability that the unit
responds to $k$ 
stimuli given the number $R$ of neurons in the unit.  Then
$\epsilon^{\rm unit}_{k,R}$ 
is given by summing Eq.\ (\ref{eq.multiunit.binomial}) over the
distributions of $\alpha_{i_r}$. In the $M$-population model for the
neurons,
\begin{align}
\label{eq.multiunit.dkR}
\epsilon^{\rm unit}_{k,R} ={}&  \int d\alpha_{i_1} \dots d\alpha_{i_R}
          { \text{Prob}(K=k \mid R, \alpha_{i_1}, \dots, \alpha_{i_R})}
\nonumber\\
  & \hspace*{2.3cm} \times \text{Prob}(\alpha_{i_1}, \dots, \alpha_{i_R})
\nonumber\\ 
={}& \sum_{i_1=1}^{M}{\dots\sum_{i_R=1}^{M}{f_{i_1}f_{i_2} \dots f_{i_R}
    \binom{S}{k}  (\alpha') ^{k} (1-\alpha') ^{S-k} }},
\end{align}
with $\alpha'$ given by Eq.\ (\ref{eq.multiunit.sparsity}).
This is in fact a version of the original multi-population model,
whose $\epsilon_k$ is given in Eq.\ (\ref{random.neuron.multi}),
with a more complicated labeling of the populations.

If the number of neurons in the unit, $R$, is sampled from a
distribution, $g(R)$, then the total probability that a unit responds
to $k$ of $S$ stimuli, $\epsilon^{\rm unit}_k$ is given by
\begin{equation}
\label{eq.multiunit.dk}
  \epsilon^{\rm unit}_k=\sum_{R=1}^\infty{g(R)\epsilon^{\rm unit}_{k,R}}.
\end{equation}
Again, this is a case of a multi-population model.
Eq.\ (\ref{eq.multiunit.dk}) is then substituted for $\epsilon_k$ in the
likelihood function, Eq.\ (\ref{likelihoodfinal}), and the rest of the
analysis proceeds in an identical fashion as with the single-neuron
unit model.

The combination of Eqs.\ (\ref{eq.multiunit.dkR}) and
(\ref{eq.multiunit.dk}) shows that from the populations at the neural
level, with their abundances and sparsities, we obtain a (larger) set
of populations at the unit level: In all cases, Eqs.\
(\ref{random.neuron.multi}), (\ref{eq.multiunit.dkR}), and
(\ref{eq.multiunit.dk}), $\epsilon_k$ is a linear combination of
binomial distributions.
Although the structure of the populations appears complicated,
considerable approximate simplifications will become apparent when we
fit data.

\section{Fits to data}
\label{sec:fits}

The data \cite{Mormann} being analyzed come from recordings of single
neuron activity in four regions of the human MTL, the hippocampus
(Hipp), the entorhinal cortex (EC), the amygdala (Amy), and the
parahippocampal cortex (PHC). Altogether, $1194$ neurons/units were
detected in the 
hippocampus, $844$ in the entorhinal cortex, $947$ in the amygdala,
and $293$ in the parahippocampal cortex, accumulated over
patients.  During the experiment, the patients were shown a randomized
sequence of images of famous individuals, landmarks, animals, and
objects. For each session, the patients were shown on average $97$
images, each of which was presented six times during the random
sequence. Histograms of the number of responses per unit for neurons
in the four MTL regions were then produced. The values are given in
Table \ref{table:rawdata}, which we read from graphs in Fig.\ 3 of
\cite{Mormann}.

\begin{table*}
  \centering
  \begin{tabular}{c|c|c|c|c|c|c|c|c|c|c|c|c|c|c|c}
   	& $n_0$ & $n_1$ & $n_2$ & $n_3$ & $n_4$ & $n_5$ & $n_6$ & $n_7$ & $n_8$ & $n_9$ & $n_{10}$ & $n_{11}$ & $n_{12}$ & $n_{13}$ & $n_{14}$ \\ \hline
    Hipp & $1019$ & $113$ & $30$ & $17$ & $7$ & $4$ & $1$ & $2$ & $0$ & $0$ & $0$ & $0$ & $1$ & $0$ & $0$ \\
    EC & $761$ & $45$ & $15$ & $9$ & $4$ & $8$ & $0$ & $0$ & $1$ & $0$ & $0$ & $0$ & $0$ & $1$ & $0$ \\
    Amy & $842$ & $61$ & $17$ & $15$ & $3$ & $3$ & $1$ & $0$ & $1$ & $1$ & $0$ & $0$ & $0$ & $1$ & $2$ \\
    PHC & $244$ & $13$ & $11$ & $7$ & $3$ & $0$ & $4$ & $1$ & $2$ & $4$ & $3$ & $0$ & $1$ & $0$ & $0$ \\ 
  \end{tabular}
  \caption{Number of neurons $n_k$ responding to $k$ images as
    reported by \cite{Mormann} in four MTL regions.} 
  \label{table:rawdata}
\end{table*}

In this section we show the results of fits to the data with the three
model implementations of our general scheme, that we summarized in
Sec.\ \ref{sec:model.defn}.  We start with a simple model
with one population of neurons with a single sparsity.  Analyzing the
disagreement between data and model leads us to first to add a
population of completely silent neurons, and then to allow the second
population a non-zero sparsity.  We fit the parameter(s) of each model
separately for each of the brain regions for which data were given.
In Table \ref{table:params}, we tabulate the numerical results of the
fits of each model in each region.  In Figs.\
\ref{fig:Hippplots}--\ref{fig:PHCplots}, we show the results
graphically; each figure shows the results of the three fits for a
single region.

\subsection{One-Population Model}

First, we attempt to fit the data in the four regions assuming only
one population of neurons with sparsity $\alpha$. In this case, Eq.\
(\ref{random.neuron.multi}) becomes: 
\begin{equation} \label{oneparam}
\epsilon_k=\binom{S}{k} {\alpha}^{k} (1-\alpha)^{S-k}.
\end{equation}
Substituting this into the likelihood function, and maximizing with
respect to $\alpha$, yields a maximum likelihood estimate (MLE) for
the sparsity.  For this and the other fits, the maximization was
performed numerically in Mathematica, and we used the exact formula
(\ref{likelihood}) for the likelihood, without any approximation such
as (\ref{likelihoodfinal}).

\begin{figure*}
\centering
\begin{tabular}{c@{\hspace*{6mm}}c}
    \includegraphics[scale=0.4]{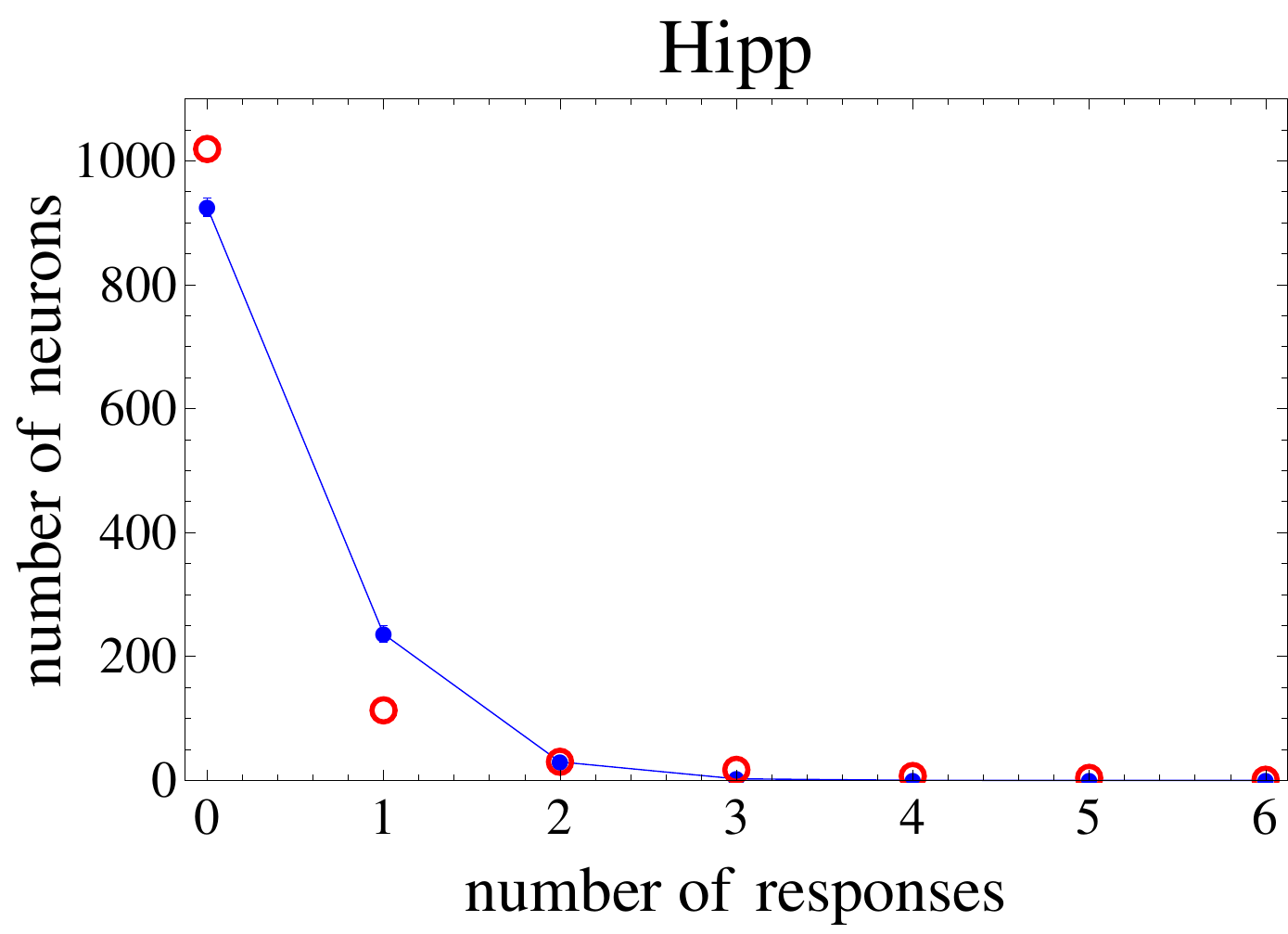} &  \includegraphics[scale=0.4]{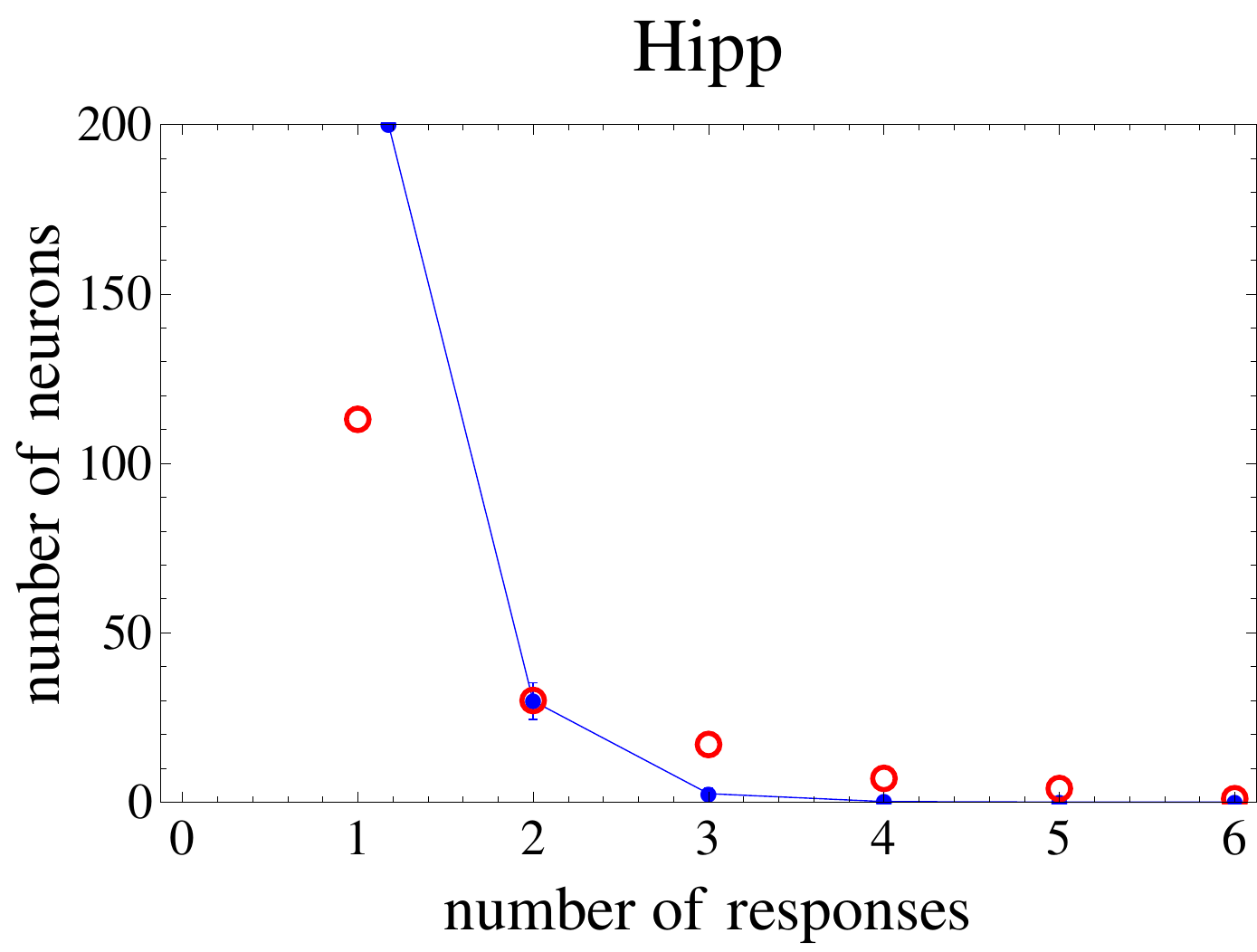} 
    \\ \includegraphics[scale=0.4]{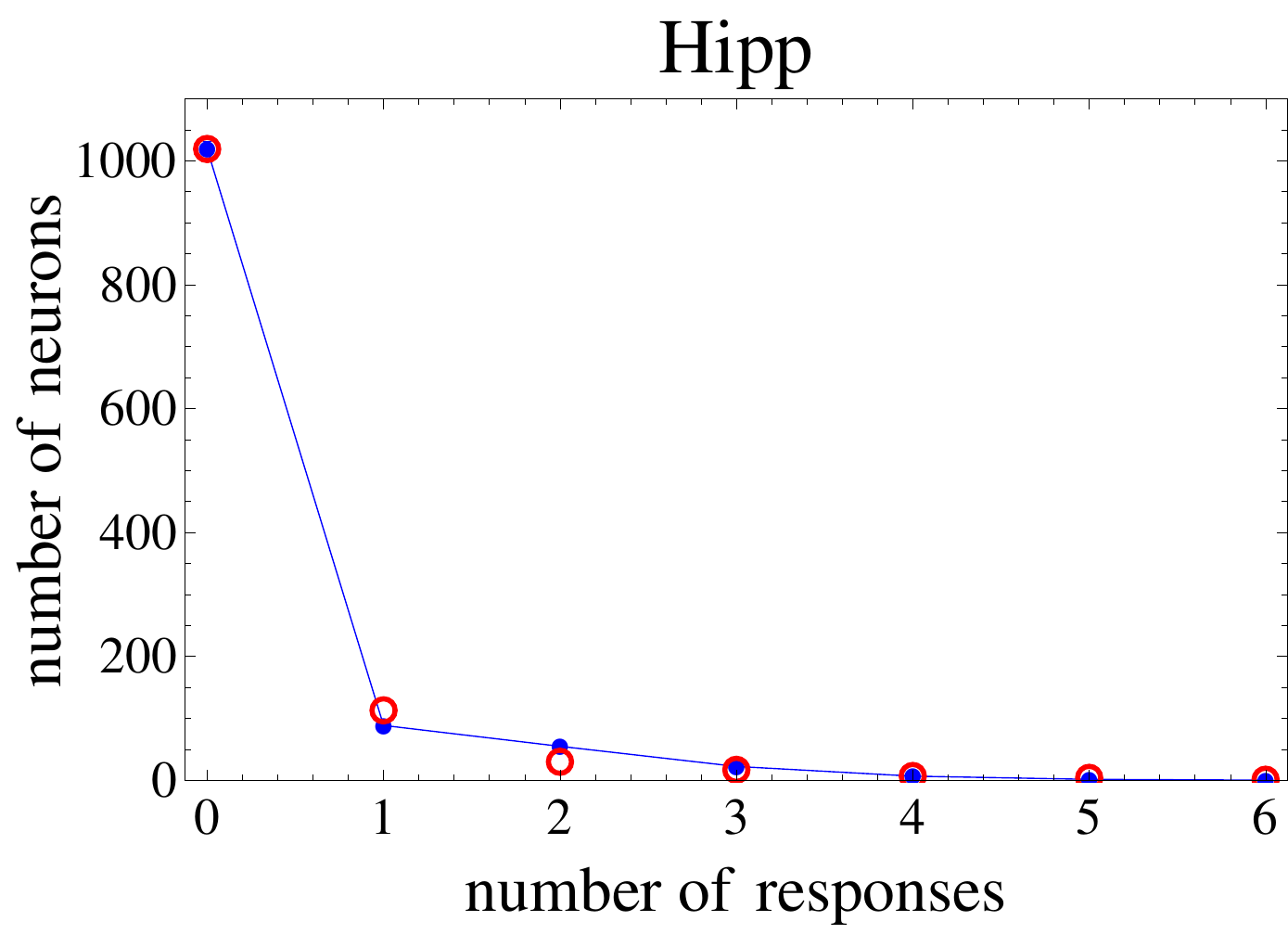} &  \includegraphics[scale=0.4]{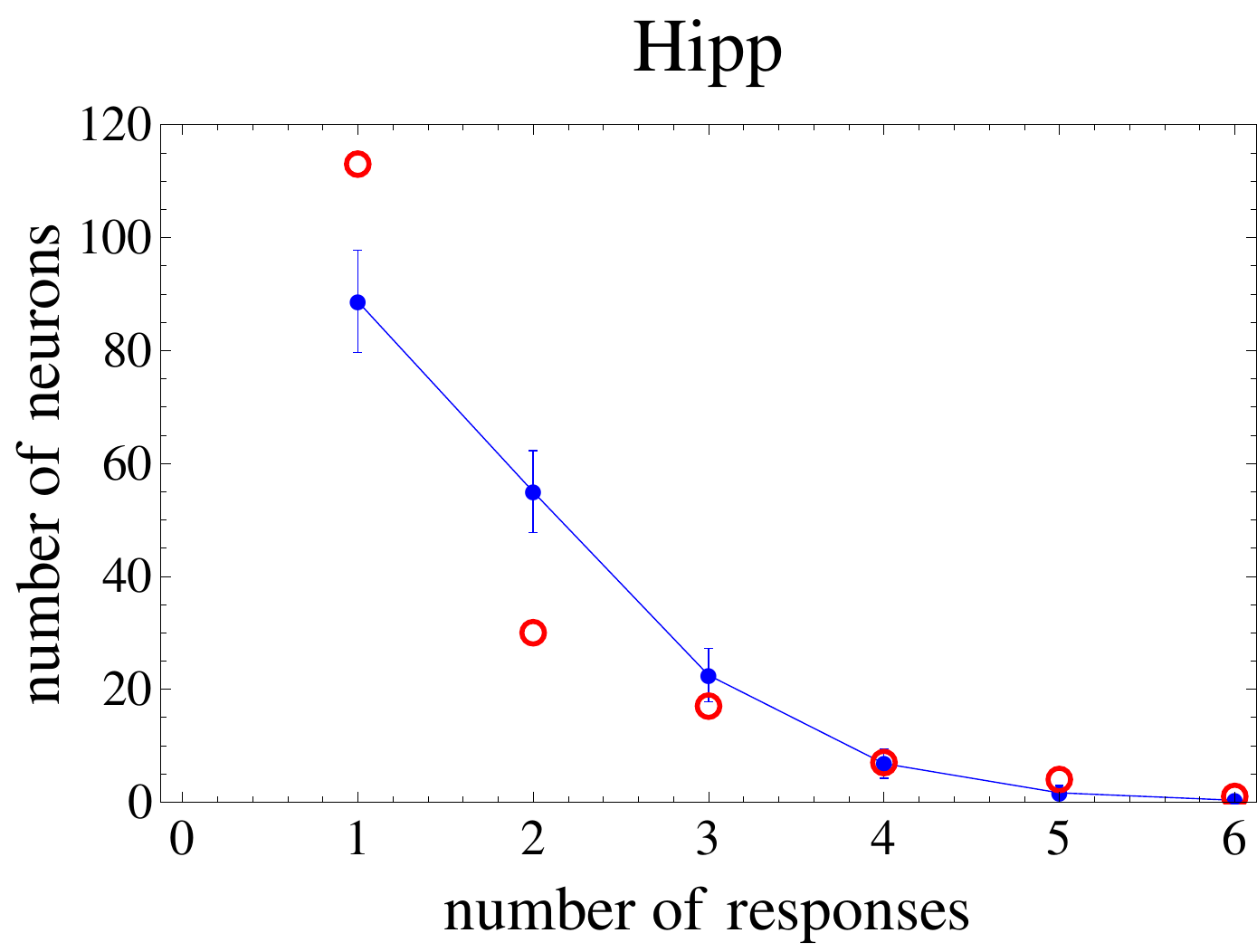} 
\\
    \includegraphics[scale=0.4]{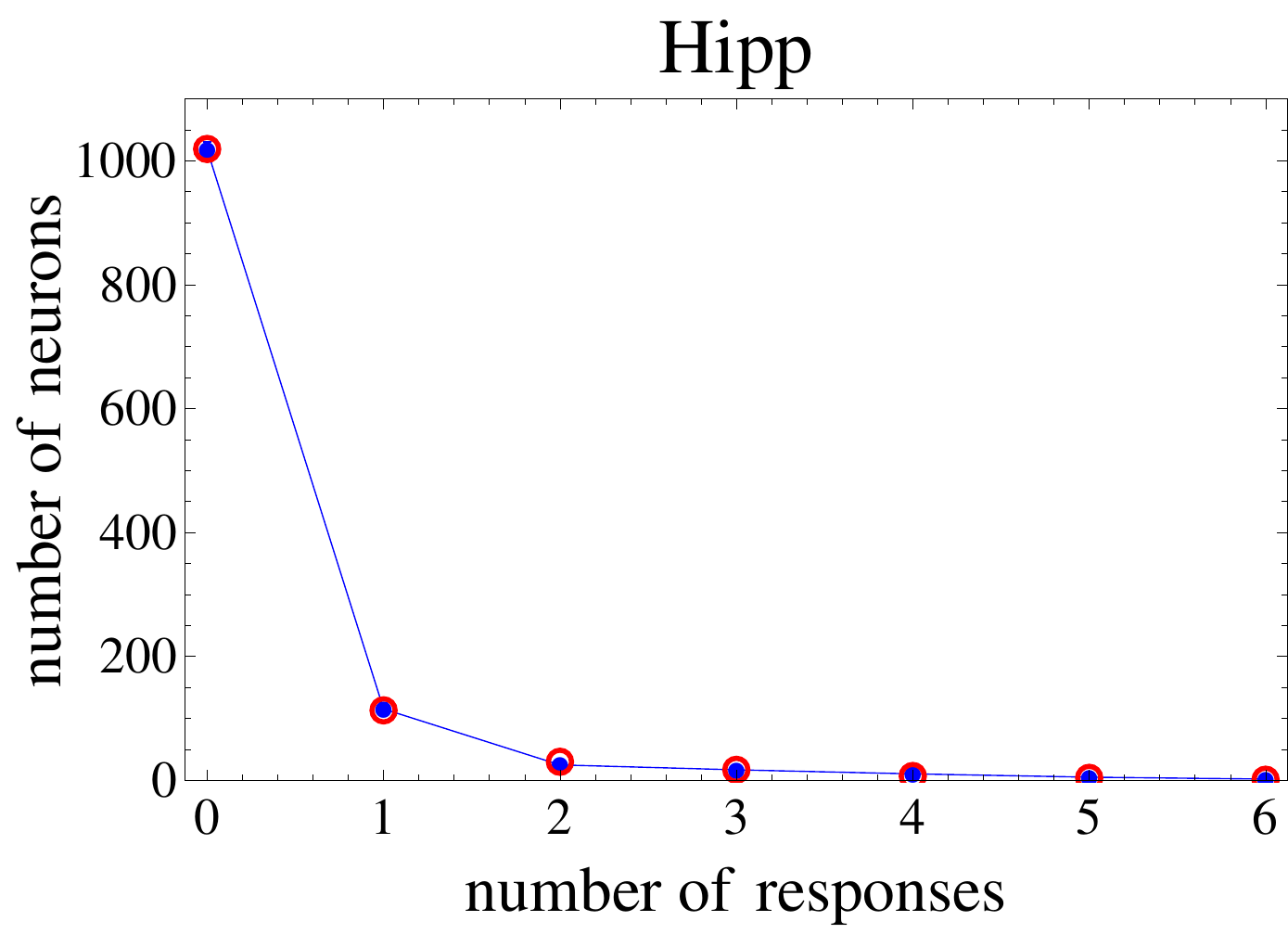} &  \includegraphics[scale=0.4]{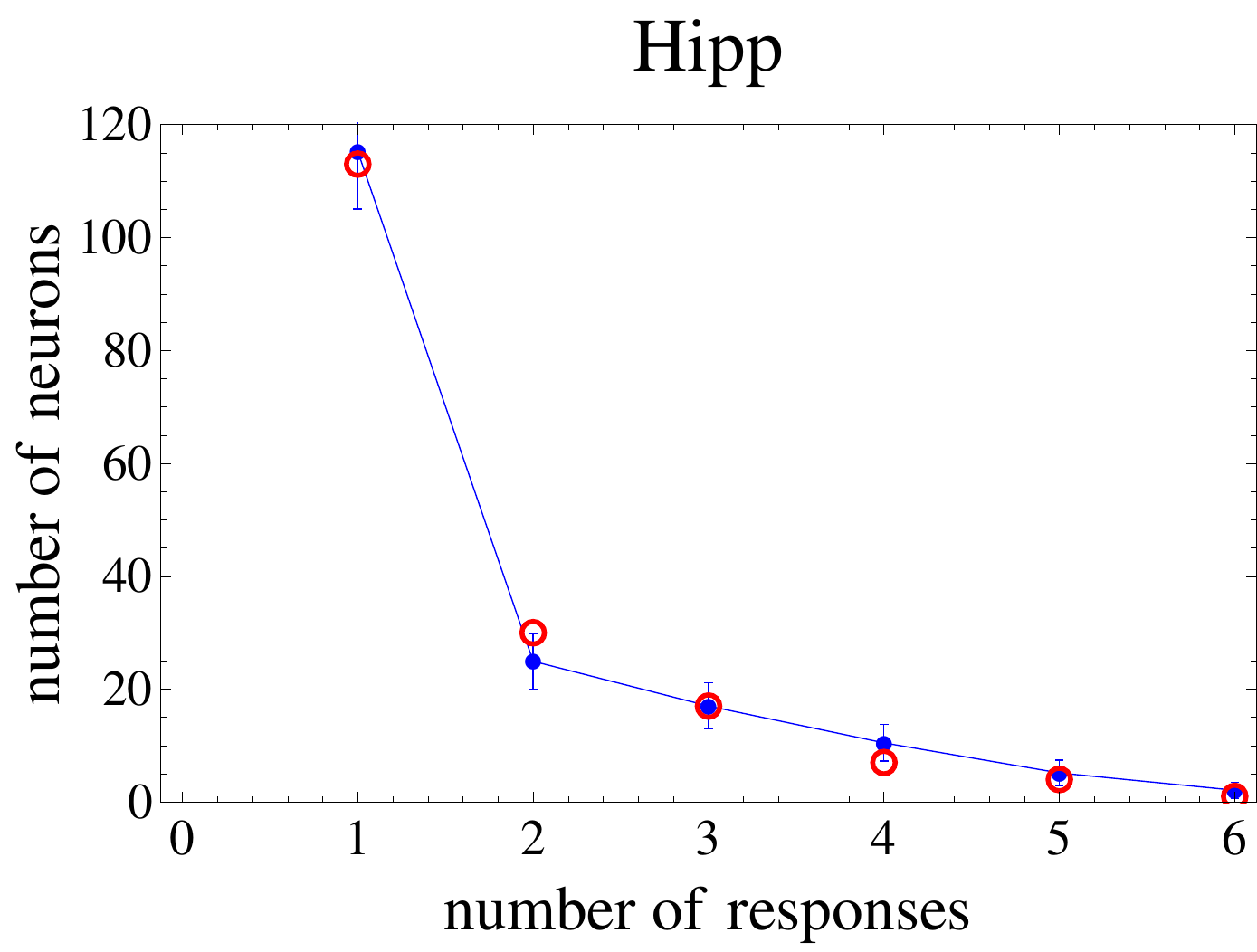}
    
\\
\multicolumn{1}{c}{(a)} & \multicolumn{1}{c}{(b)}
\end{tabular}
\caption{Comparison of data from the hippocampus with fits for the
  number of neurons $n_k$ that respond to $k$ stimuli, for each of the
  three models.  The red circles indicate the experimental values and
  blue dots connected by lines indicate the model predictions for the
  expectation values of $n_k$.  The blue error bars indicate the
  model's prediction for the one-standard-deviation variation of
  experimental results on repetition of the experiment. The top plots
  are the fits from the one-population model. The middle plots are
  fits from the model with one active and one silent population. The
  bottom fits are for the model with two active populations.  The
  left-hand plots are with the zero-response bin included, and the
  right-hand plots are without them, to show more clearly the other
  bins.}
\label{fig:Hippplots}
\end{figure*}

\begin{figure*}
\centering
\begin{tabular}{c@{\hspace*{6mm}}c}
    \includegraphics[scale=0.4]{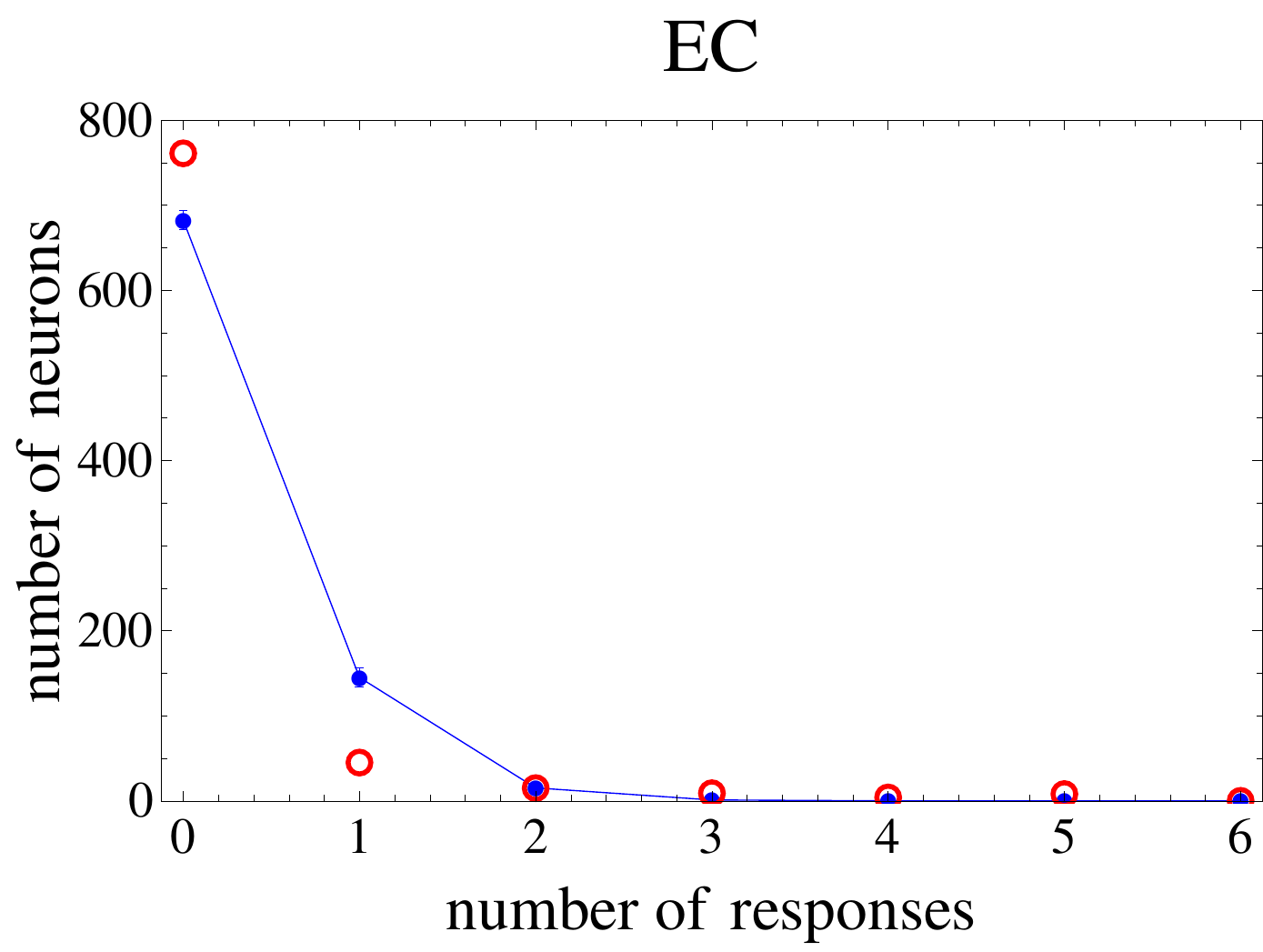} &  \includegraphics[scale=0.4]{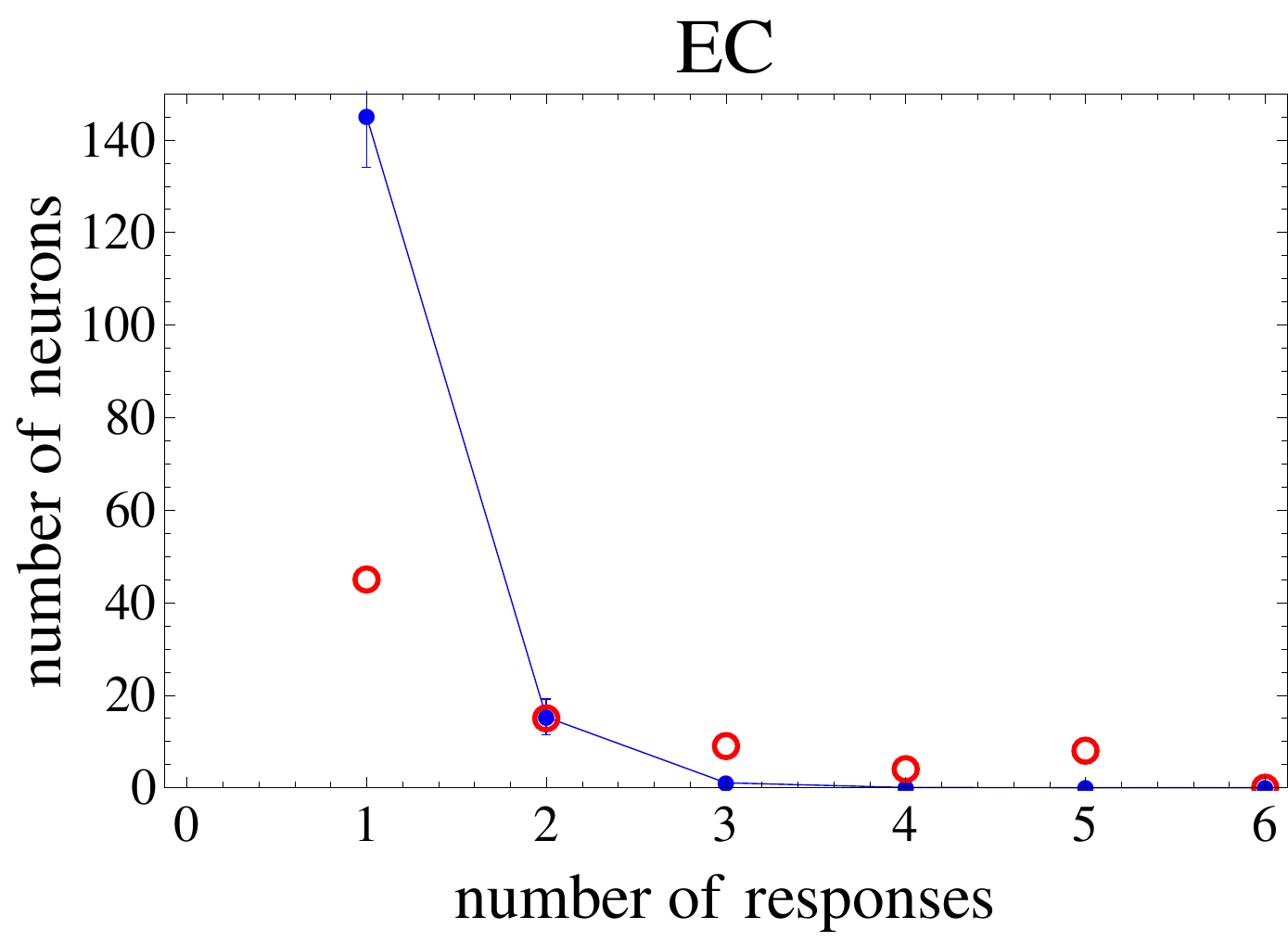} 
    \\ \includegraphics[scale=0.4]{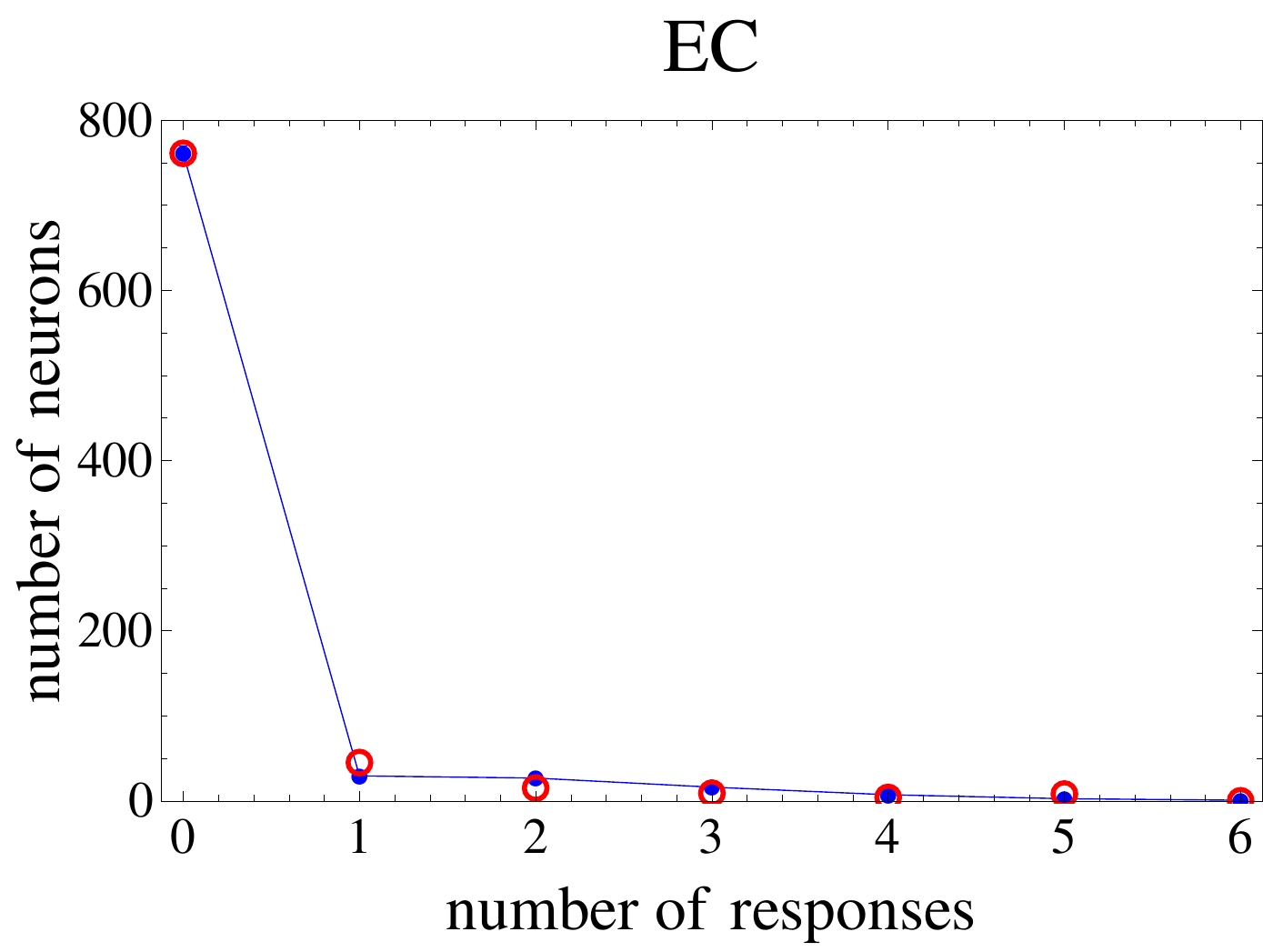} &  \includegraphics[scale=0.4]{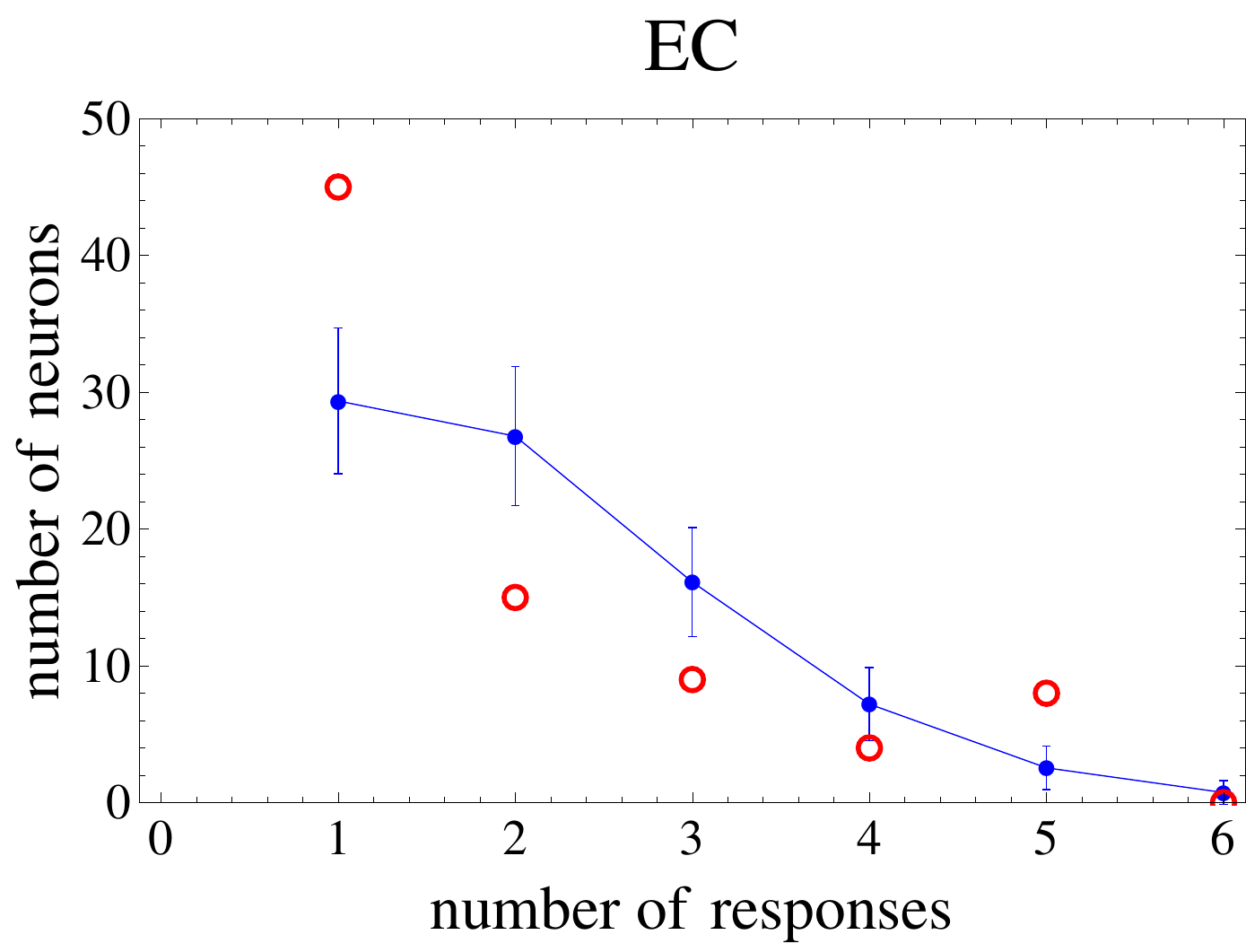} 
\\
    \includegraphics[scale=0.4]{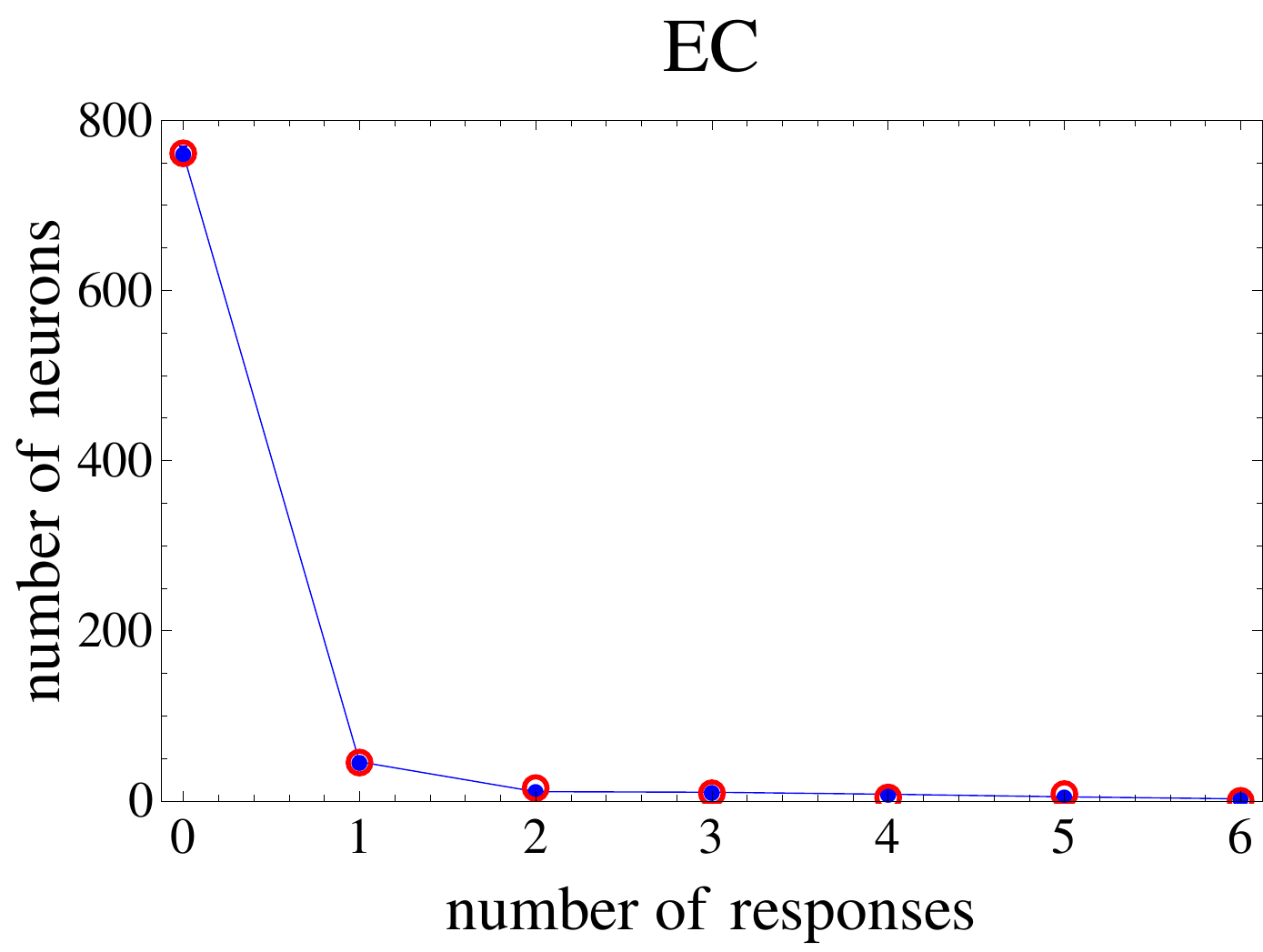} &  \includegraphics[scale=0.4]{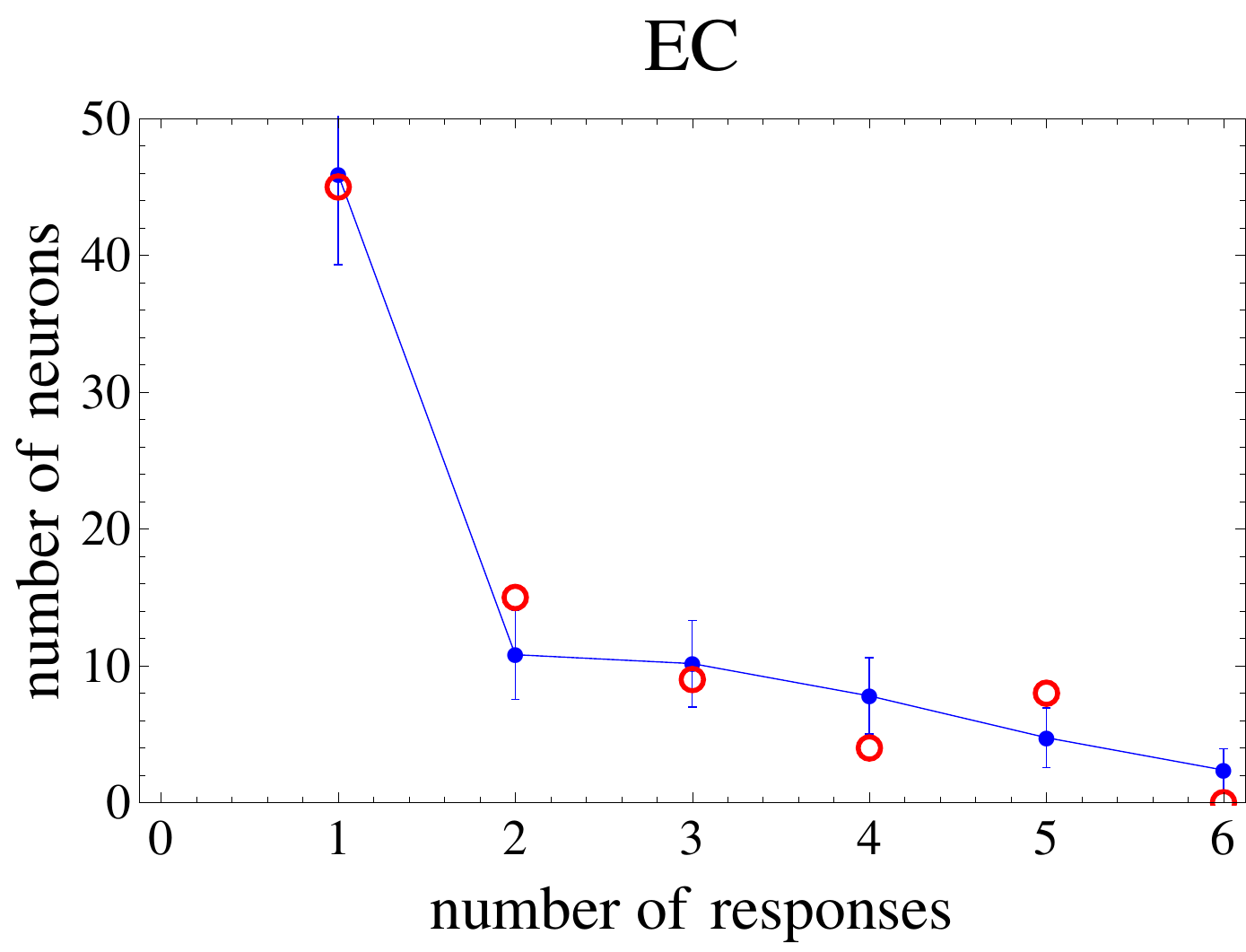}
    
\\
\multicolumn{1}{c}{(a)} & \multicolumn{1}{c}{(b)}
\end{tabular}
\caption{The same as Fig.\ \ref{fig:Hippplots}, but for the entorhinal
  cortex.}
\label{fig:ECPlots}
\end{figure*}

\begin{figure*}
\centering
\begin{tabular}{c@{\hspace*{6mm}}c}
    \includegraphics[scale=0.4]{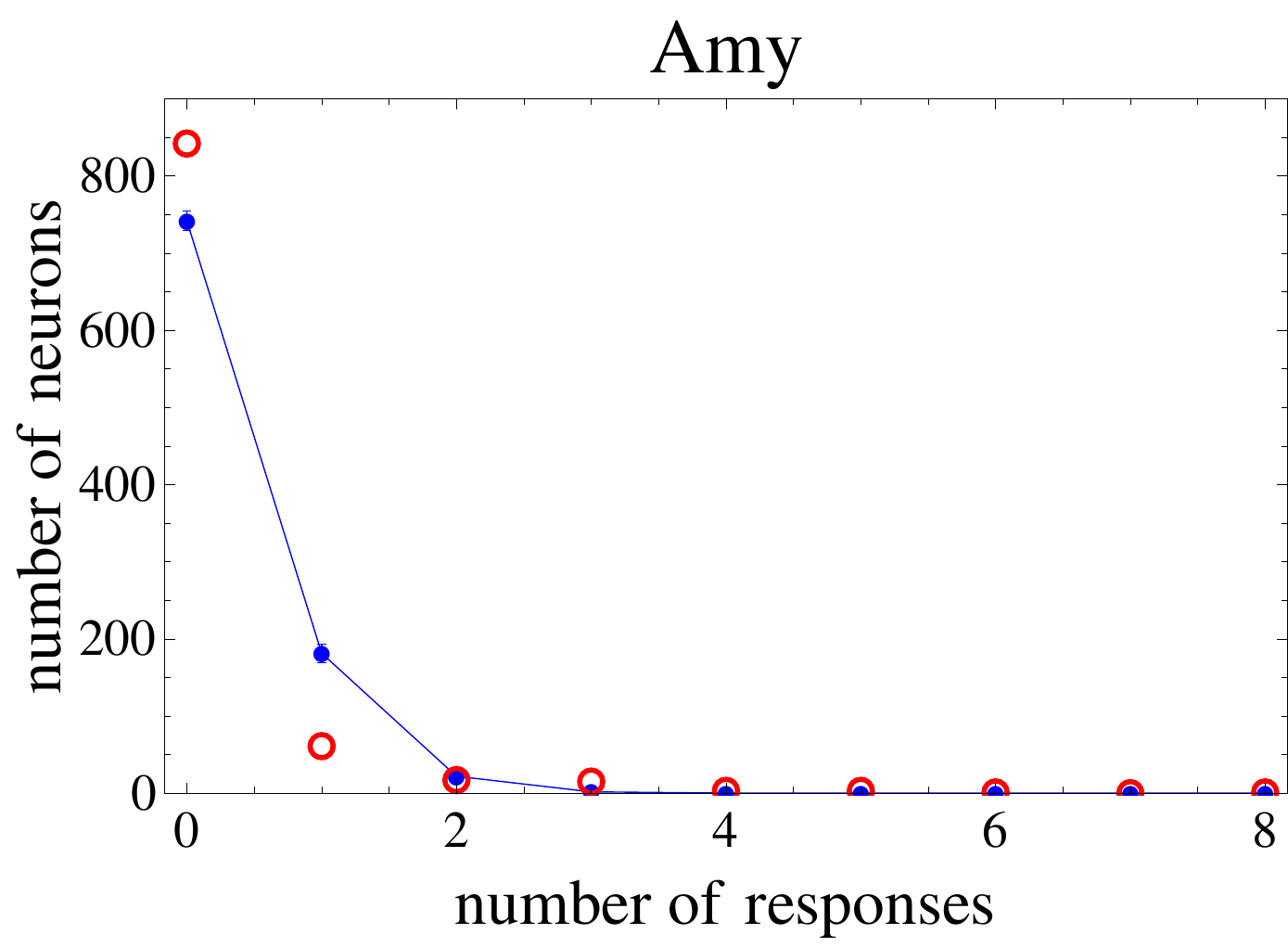} &  \includegraphics[scale=0.4]{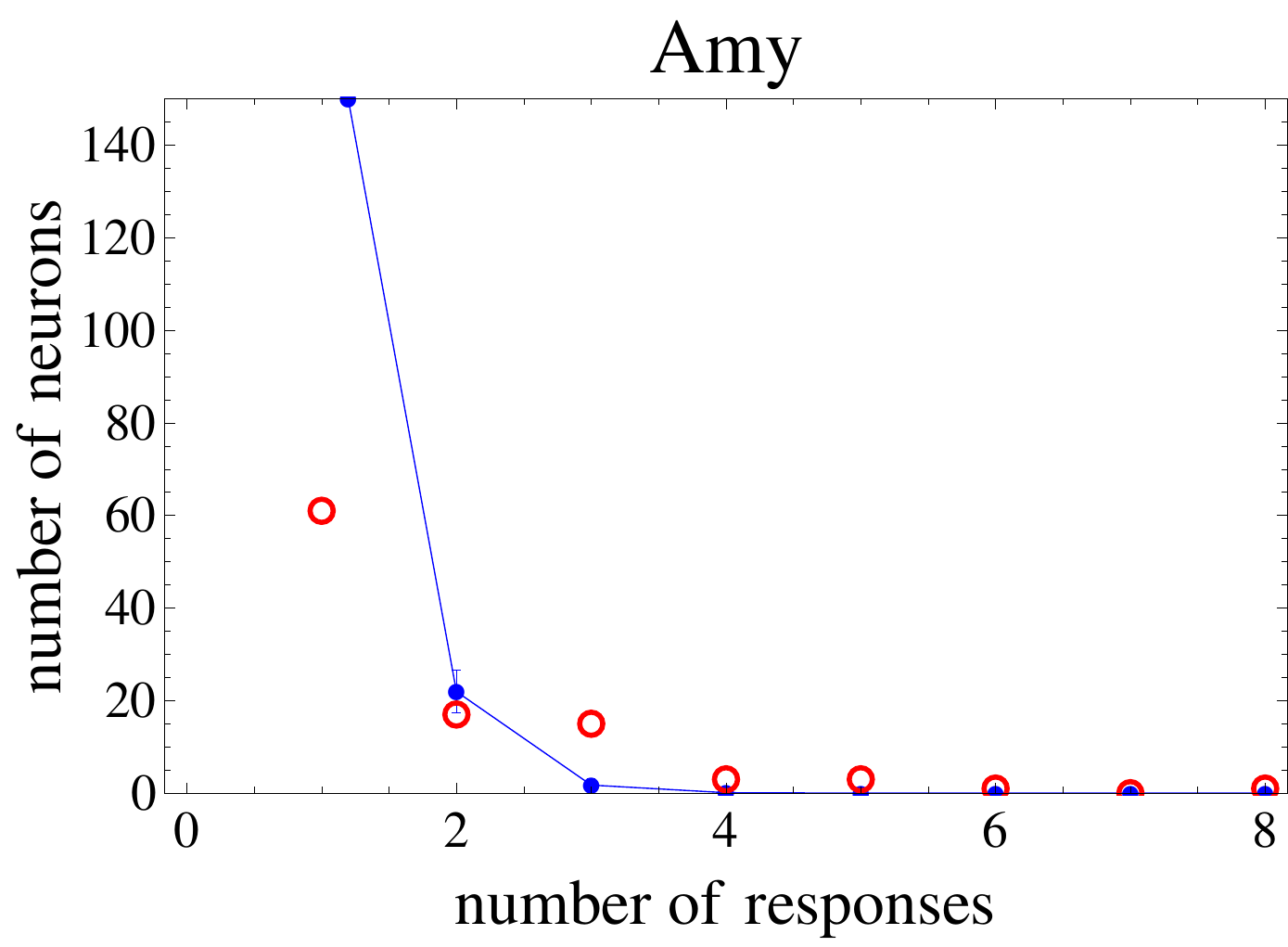} 
    \\ \includegraphics[scale=0.4]{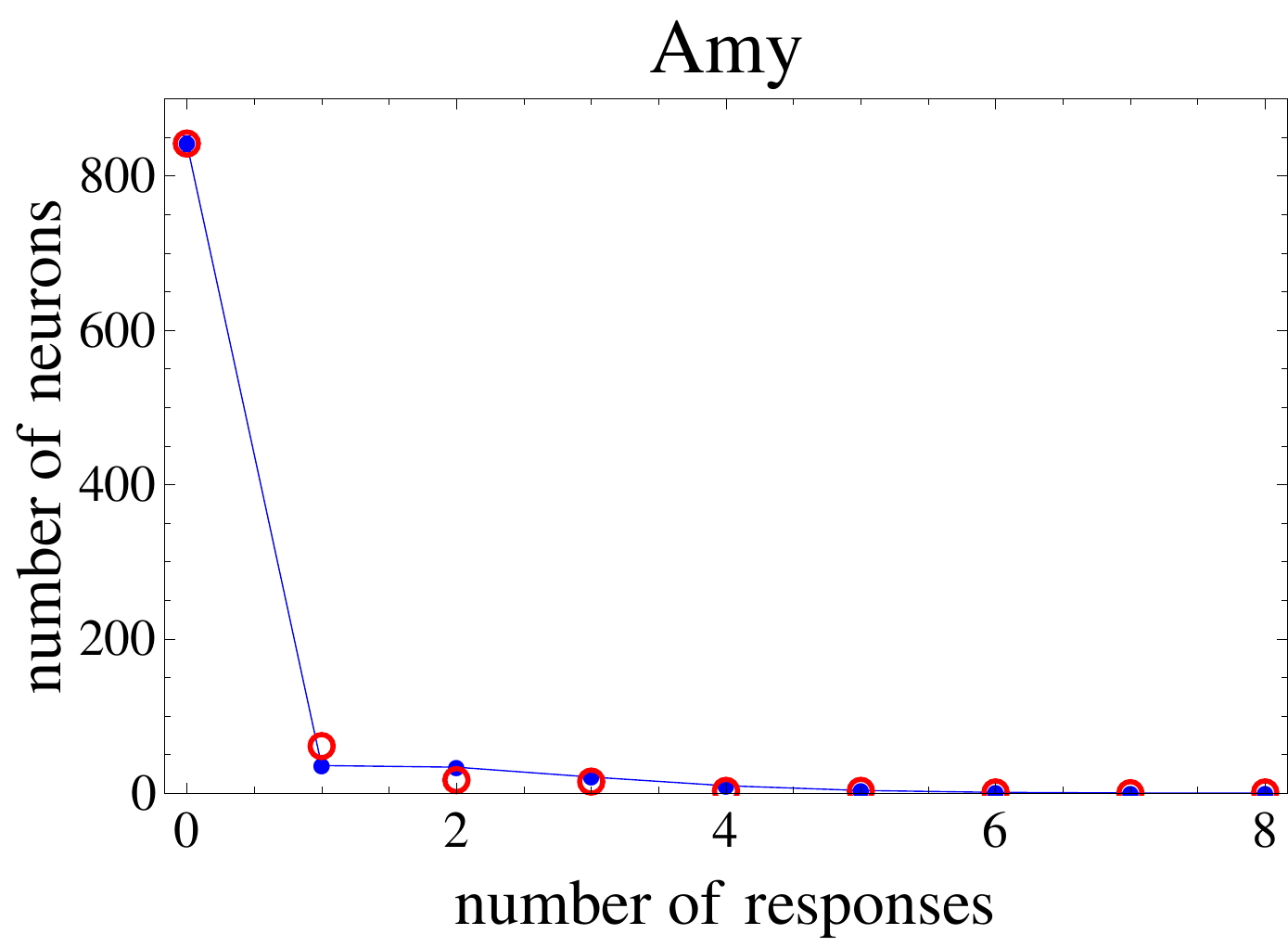} &  \includegraphics[scale=0.4]{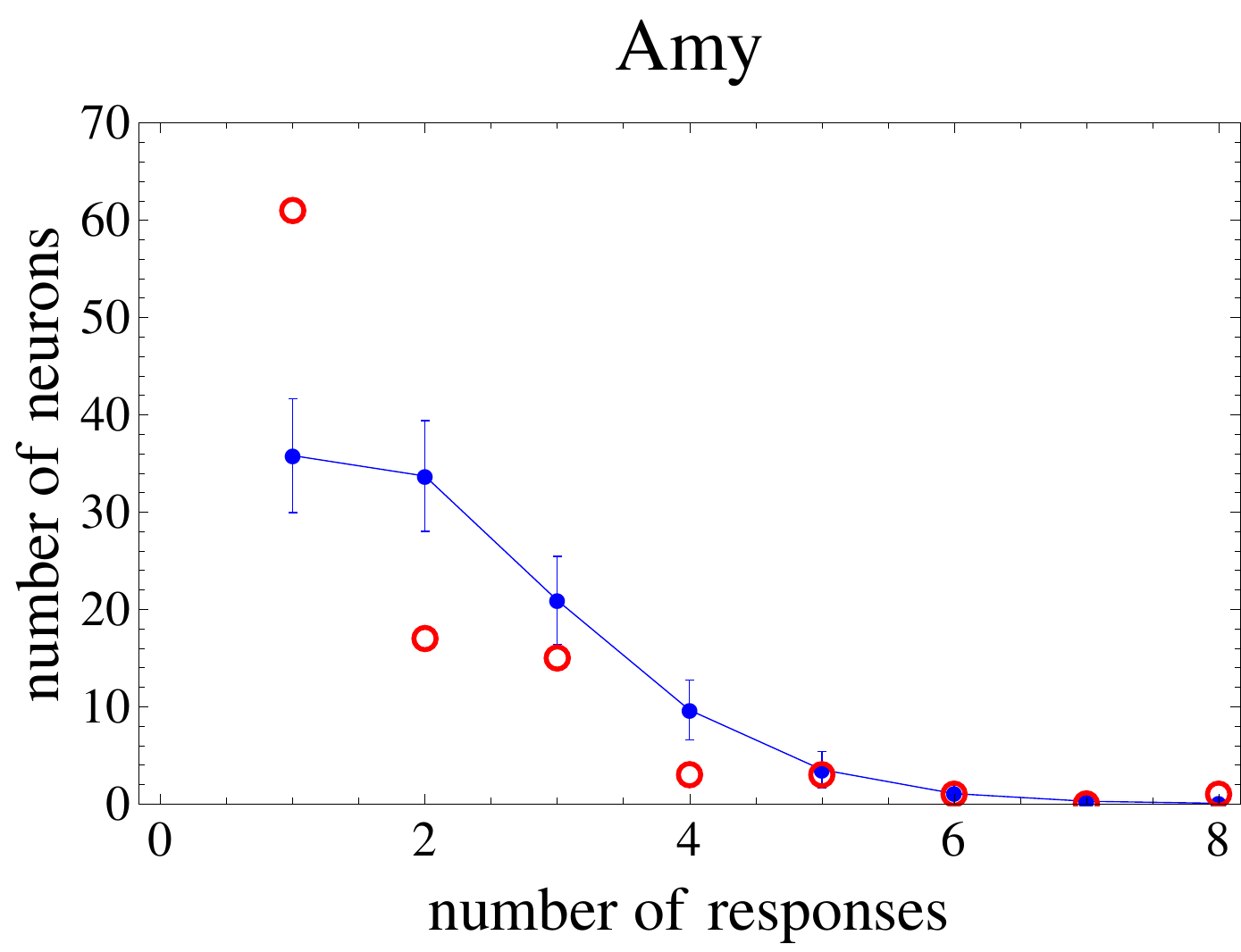} 
\\
    \includegraphics[scale=0.4]{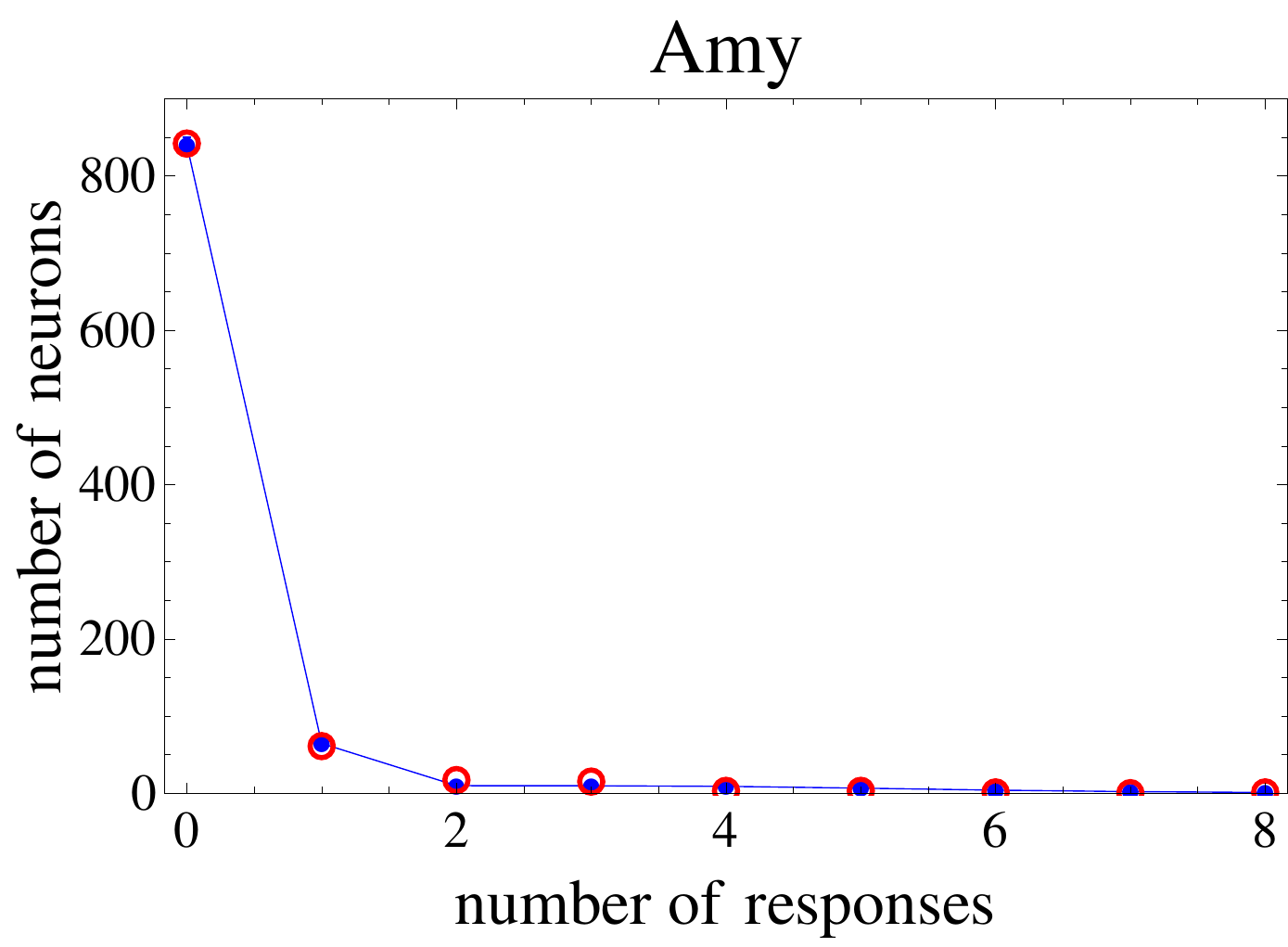} &  \includegraphics[scale=0.4]{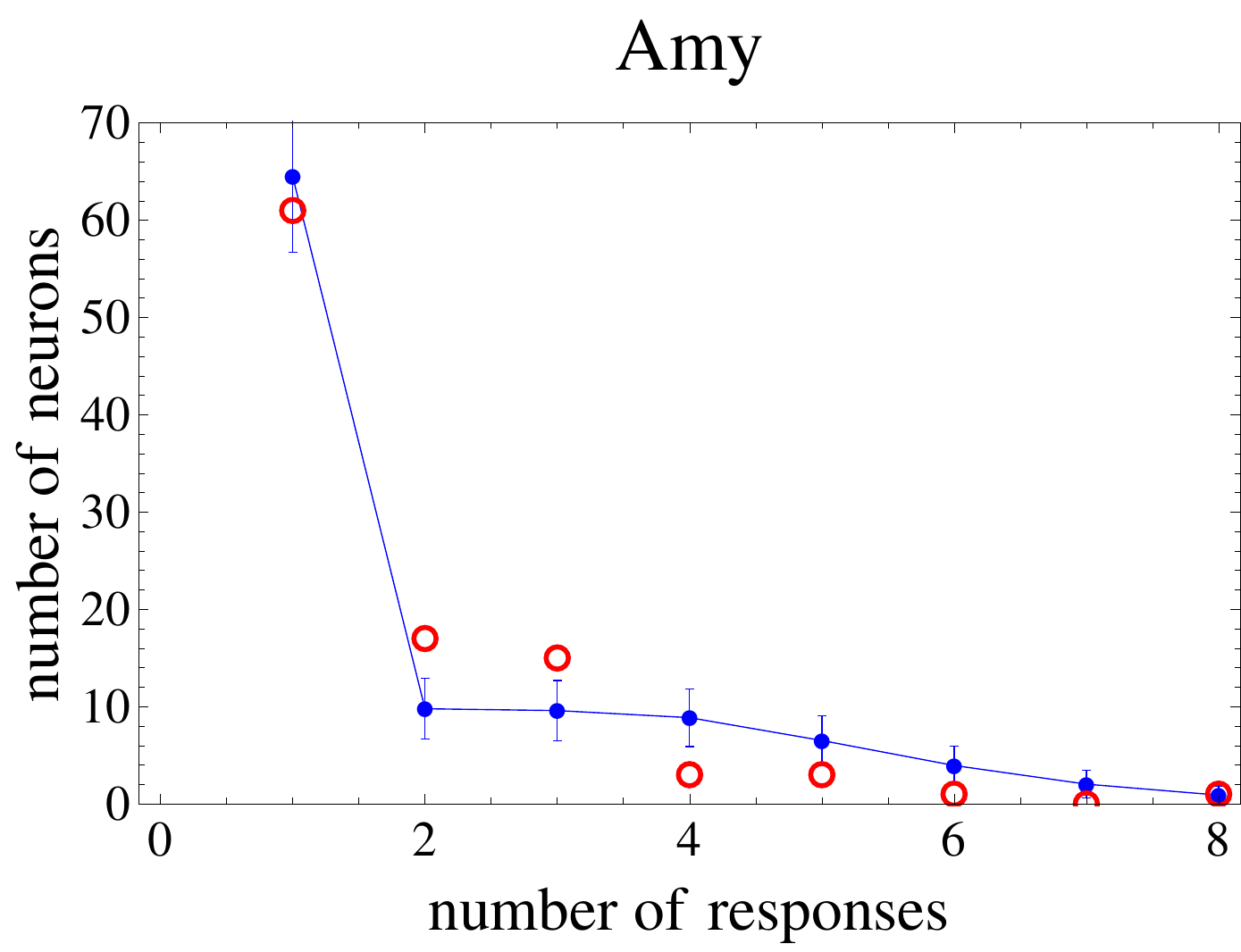}
    
\\
\multicolumn{1}{c}{(a)} & \multicolumn{1}{c}{(b)}
\end{tabular}
\caption{The same as Fig.\ \ref{fig:Hippplots}, but for the amygdala.}
\label{fig:Amyplots}
\end{figure*}

\begin{figure*}
\centering
\begin{tabular}{c@{\hspace*{6mm}}c}
    \includegraphics[scale=0.4]{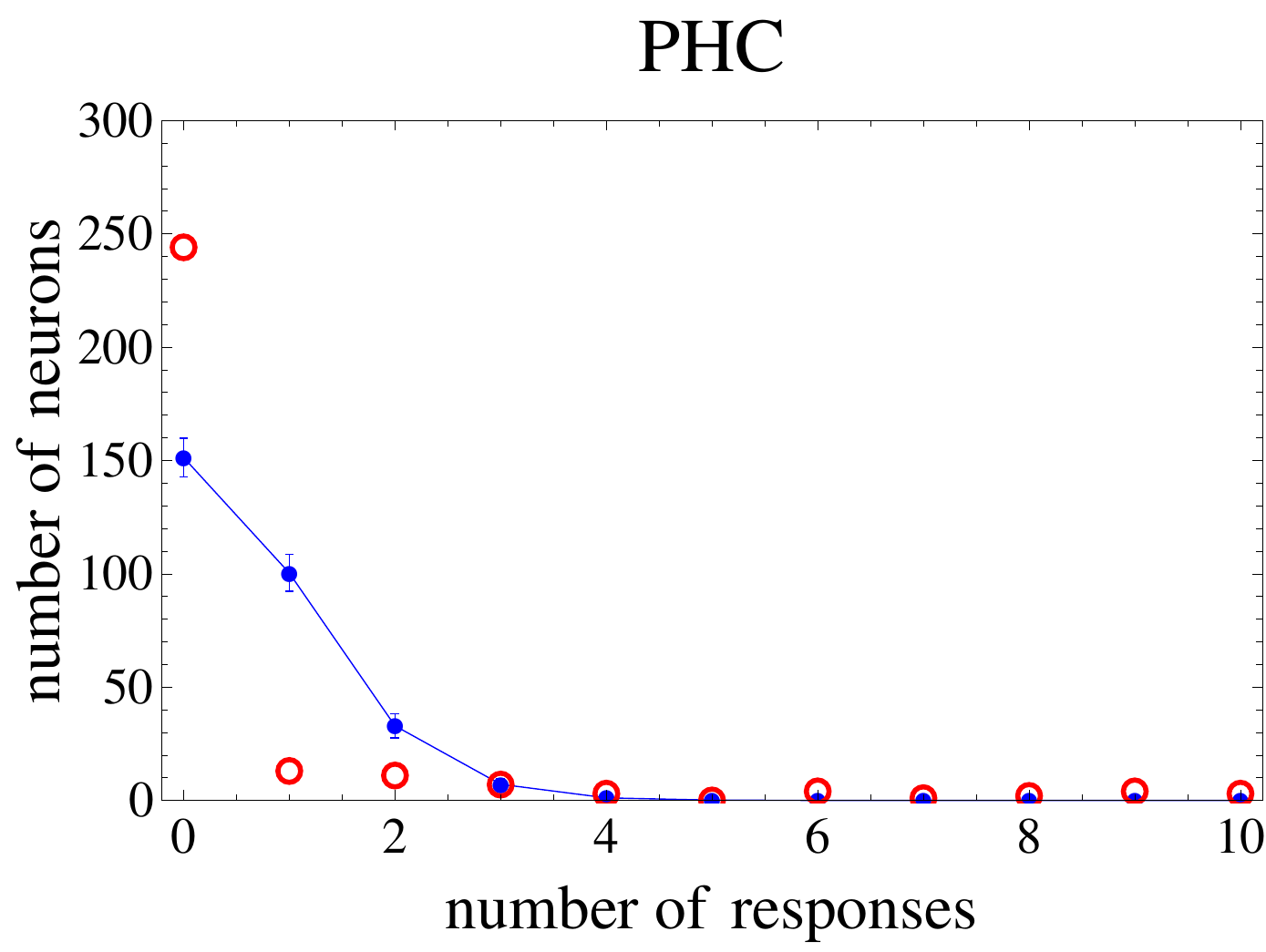} &  \includegraphics[scale=0.4]{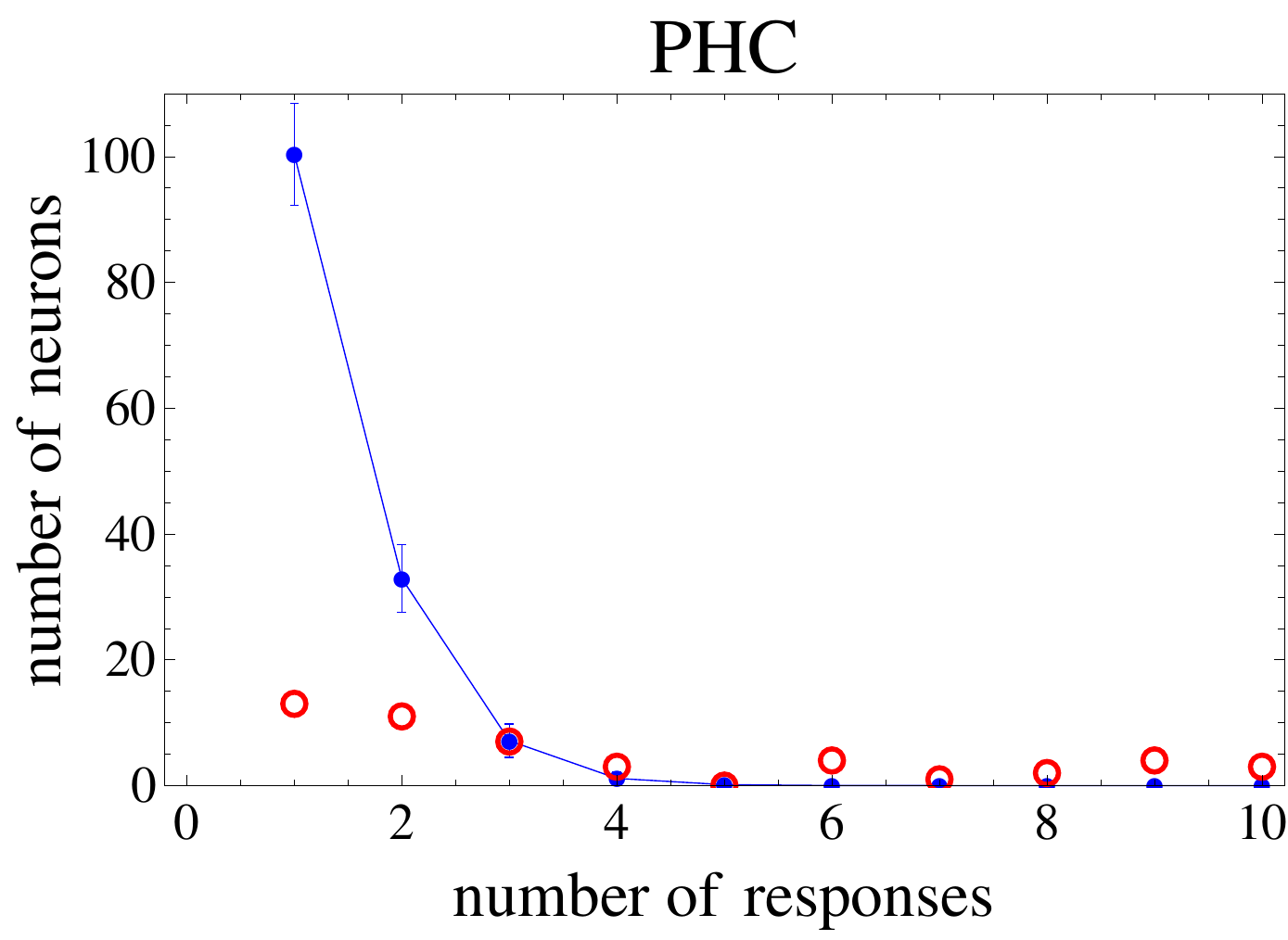} 
    \\ \includegraphics[scale=0.4]{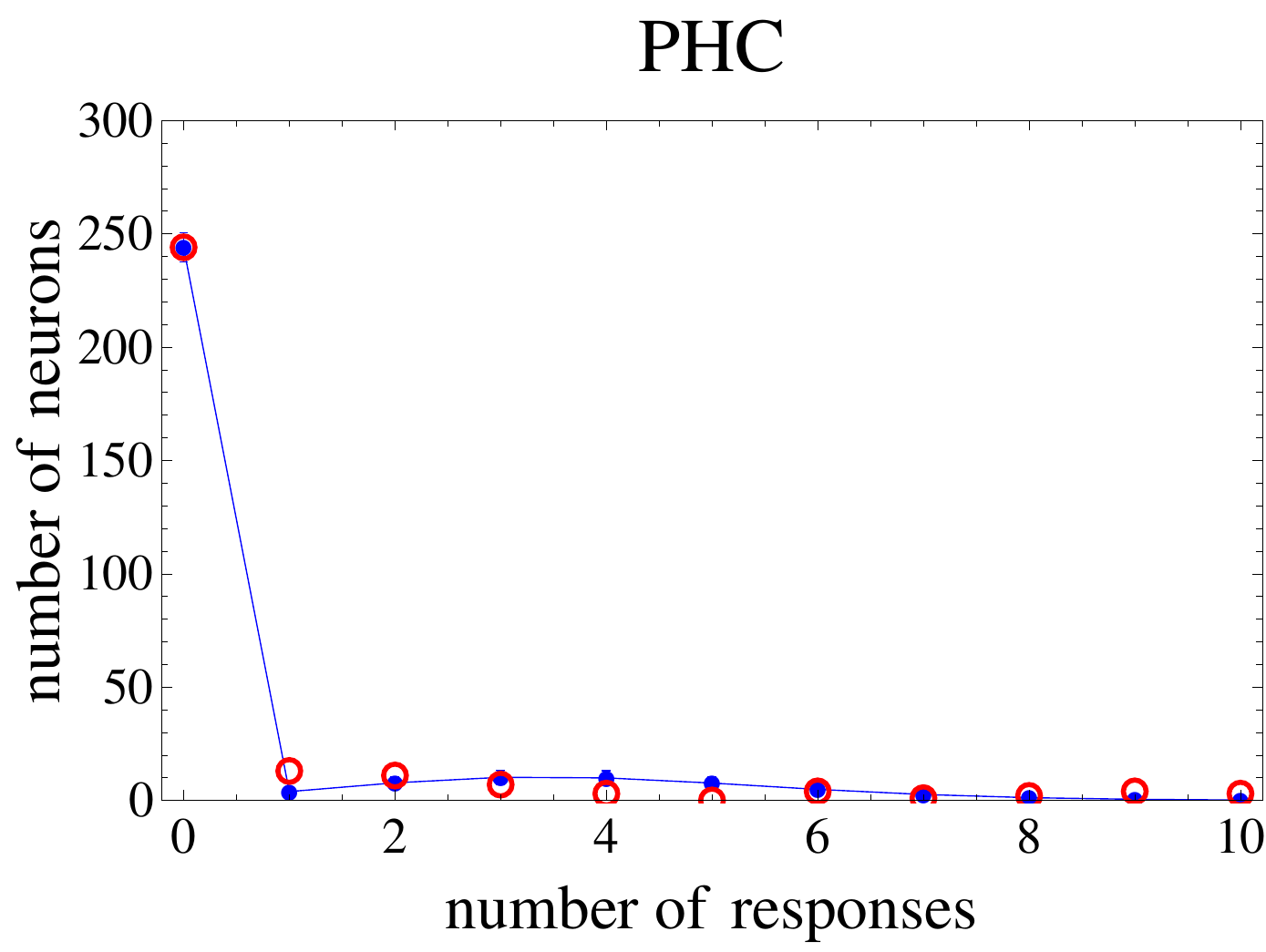} &  \includegraphics[scale=0.4]{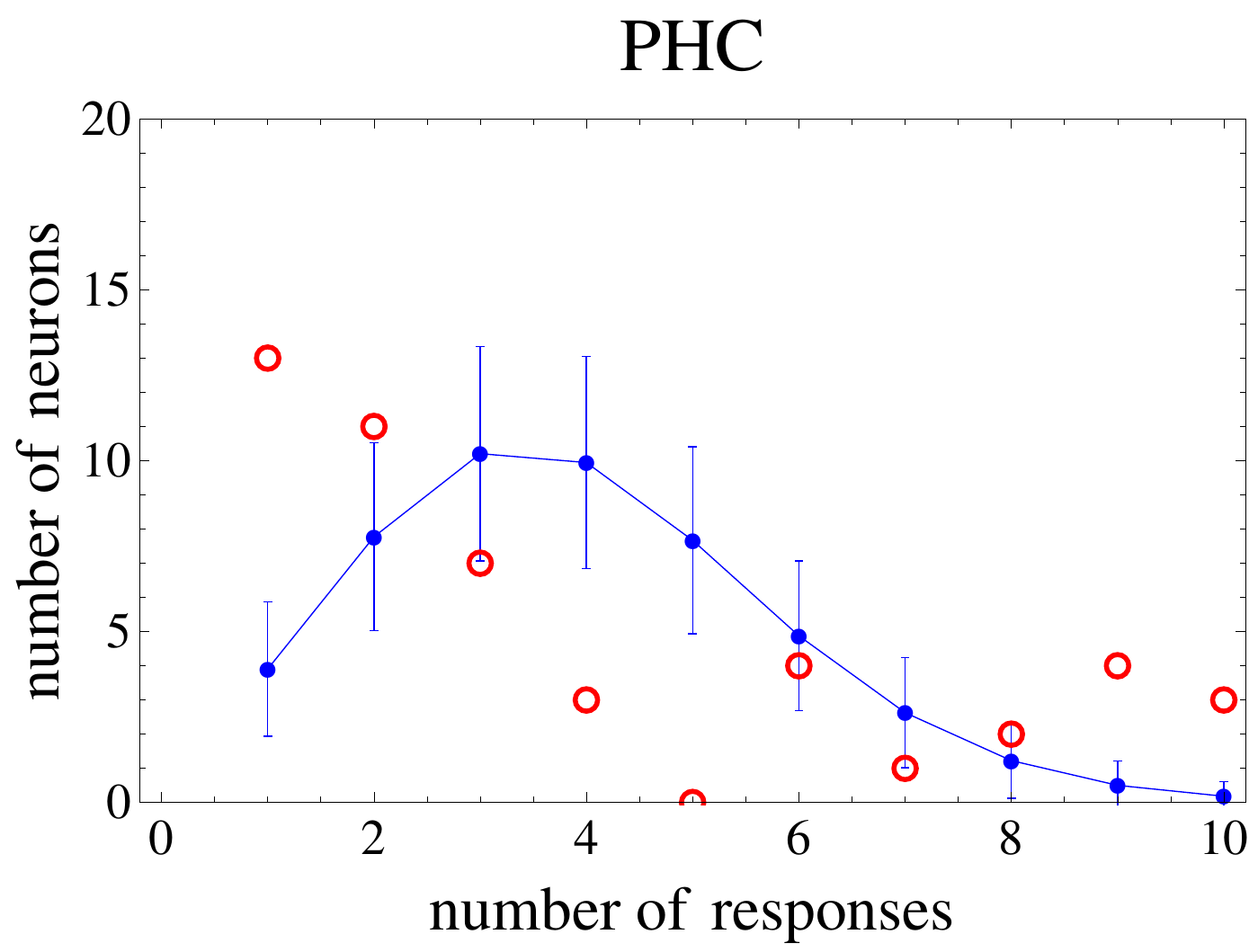} 
\\
    \includegraphics[scale=0.4]{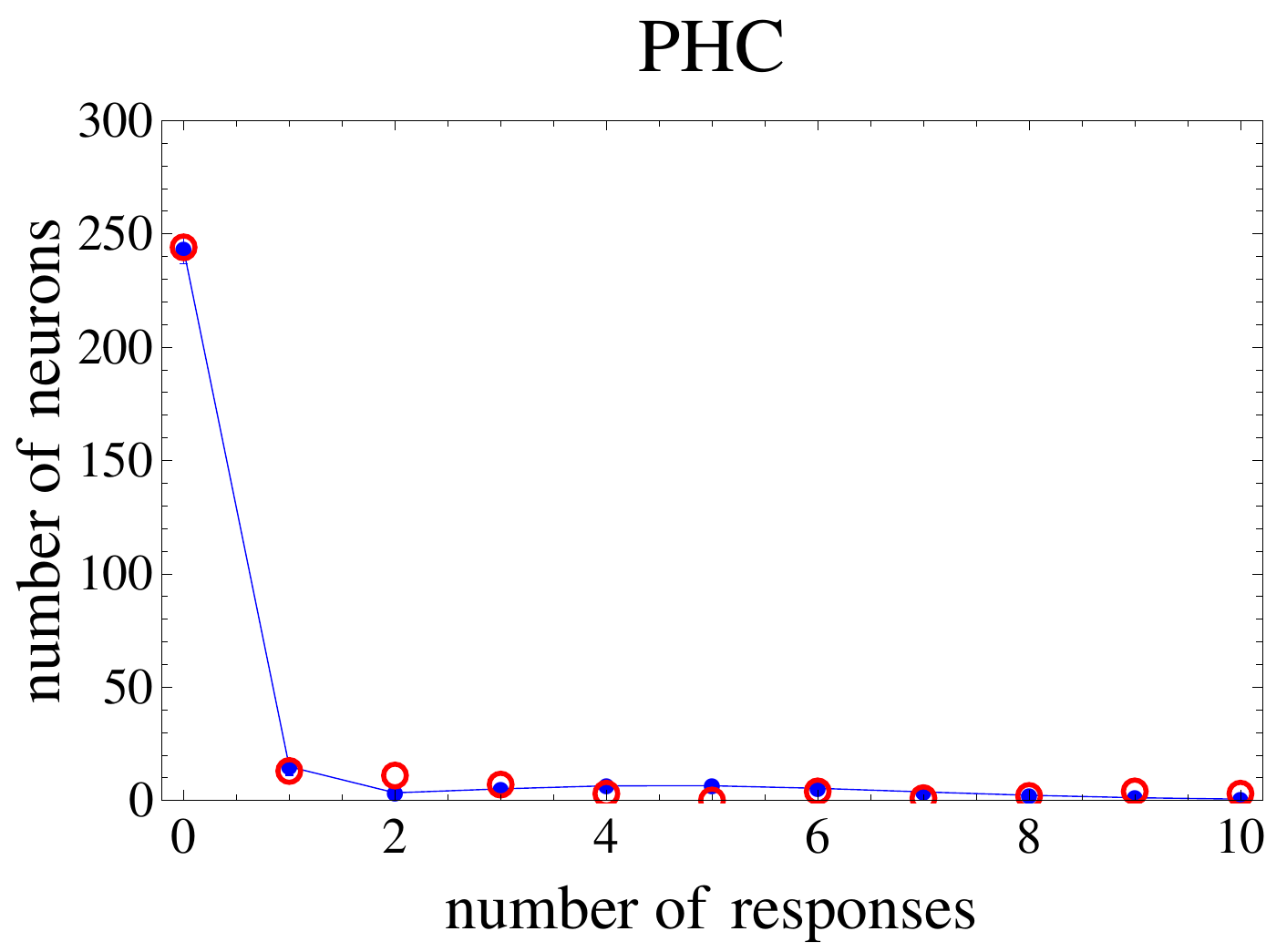} &  \includegraphics[scale=0.4]{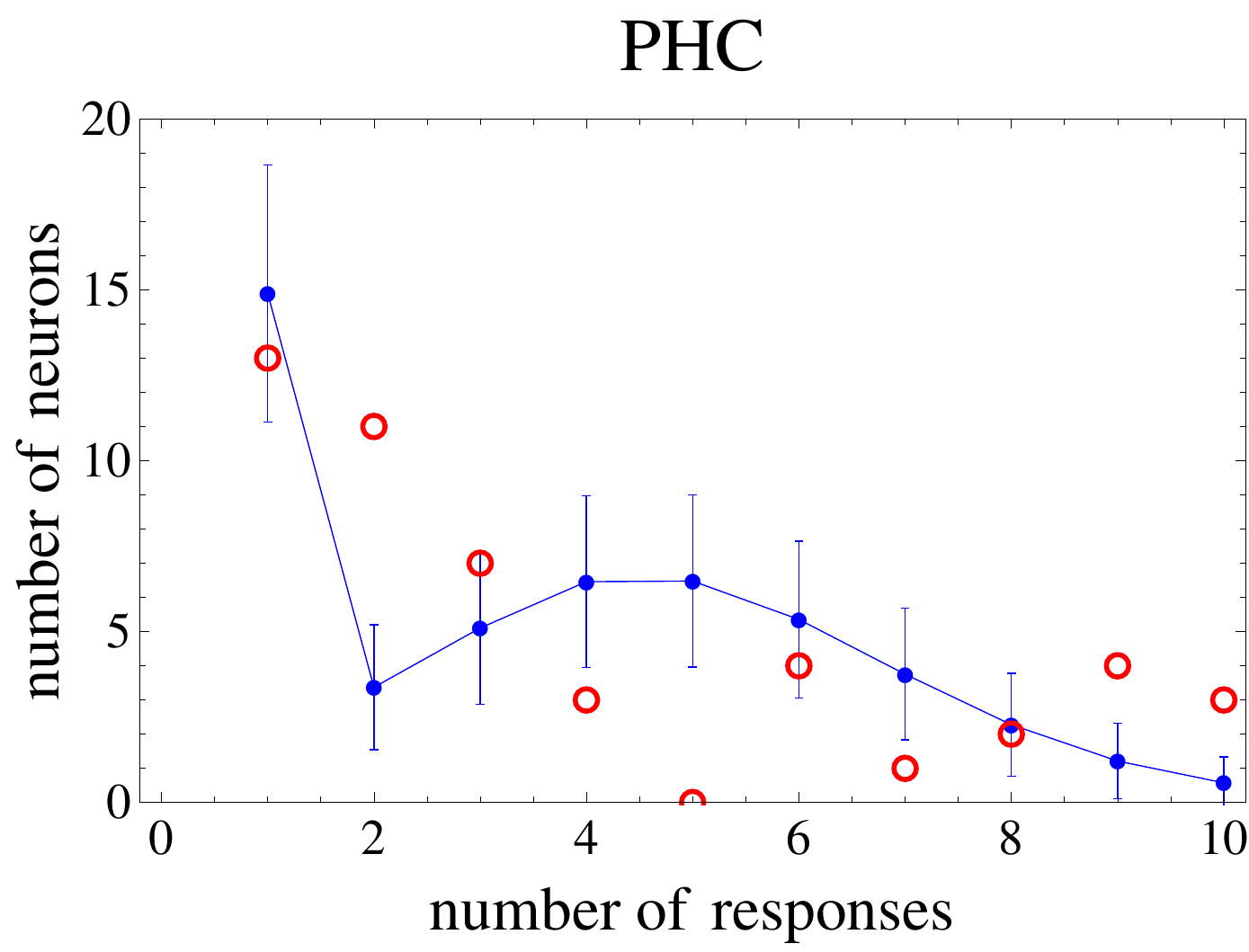}
    
\\
\multicolumn{1}{c}{(a)} & \multicolumn{1}{c}{(b)}
\end{tabular}
\caption{The same as Fig.\ \ref{fig:Hippplots}, but for the parahippocampal cortex.}
\label{fig:PHCplots}
\end{figure*}

The parameters of the fits are shown numerically in Table
\ref{table:params}(a).  In the top row of each of Figs.\
\ref{fig:Hippplots}--\ref{fig:PHCplots}, the data for each region are
compared with the expectation values of the response counts $N_k$
\begin{equation}\label{prediction}
  \langle N_k \rangle =N\epsilon_k.
\end{equation}
The error bars in the plots are the one-standard-deviation variations
that the model predicts for the $N_k$ over repetitions of the whole
experiment.

\begin{table*}
  \centering
  \begin{tabular}{c}
    \begin{tabular}{c|c|c|c|c}
                    & Hipp                      & EC                        & Amy                        & PHC                       \\ \hline
      $\alpha$      & $(2.6\pm0.1)\times10^{-3}$ & $(2.2\pm0.2)\times10^{-3}$ & $(2.5\pm0.2)\times10^{-3}$ & $(6.8\pm0.5)\times10^{-3}$ \\ \hline
      $\chi^2(5)$   & $5.1\times10^2 $          & $4.8\times10^2$            & $3.4\times10^2 $          & $2.5\times10^2$                       \\
    \end{tabular}
    \vspace*{3mm}
    \\
      (a) One-population model.
  \\*[8mm]
    \begin{tabular}{c|c|c|c|c}
                        & Hipp                        & EC                        & Amy                         & PHC                        \\ \hline
      $\alpha_{\rm D}$   & $(1.3\pm0.1)\times10^{-2}$ & $(1.9\pm0.2)\times10^{-2}$  & $(1.9\pm0.1)\times10^{-2}$  & $(4.0\pm0.3)\times10^{-2}$  \\ \hline
      $f_{\rm D}$        & $0.21\pm0.02$               & $0.11\pm0.01$             & $0.13\pm0.01$               & $0.17\pm0.02$              \\ \hline
      $\chi^2(5)$       & $20$                        & $19$                      & $33$                        & $29$                       \\ 
    \end{tabular}
    \vspace*{3mm}
    \\
    (b) Two-population model, with one population being totally
        silent. 
  \\*[8mm]
    \begin{tabular}{c|c|c|c|c}
                        & Hipp                       & EC                          & Amy                         & PHC                        \\ \hline
      $\alpha_{\rm D}$   & $(2.6\pm0.3)\times10^{-2}$  & $(3.2\pm0.4)\times10^{-2}$  & $(3.8\pm0.4)\times10^{-2}$   & $(5.1\pm0.4)\times10^{-2}$  \\ \hline
      $f_{\rm D}$        & $0.06\pm0.01$             & $0.05\pm0.01$              & $0.05\pm0.01$              & $0.12\pm0.02$              \\ \hline
      $\alpha_{\rm US}$  & $(1.0\pm0.1)\times10^{-3}$  & $(5.4\pm1.0)\times10^{-4}$  & $(7.4\pm1.1)\times10^{-4}$   & $(5.8\pm2.0)\times10^{-4}$  \\ \hline
      $f_{\rm US}$       & $0.94\pm0.01$              & $0.95\pm0.01$               & $0.95\pm0.01$               & $0.88\pm0.02$              \\ \hline
      $\chi^2(5)$       & $2.3$                      & $3.7$                       & $14.3$                      & $20$                       \\ \hline
      $\chi^2(10)$      & $5.7$                      & $10.5$                      & $19.9$                      & $36$                       \\
    \end{tabular}
    \vspace*{3mm}
    \\
    (c) Full two-population model.
  \end{tabular}
  \caption{Results of the fits of the three models.  For each model,
    we give the values and uncertainties of the model's parameters for
    each brain region, and we give the $\chi^2$ measuring the goodness
    of fit.}
  \label{table:params}
\end{table*}

We assess the goodness of fit of the model to the data by the $\chi^2$
test, with $\chi^2$ defined in Eq.\ (\ref{eq:chi2}). The enormous
values of $\chi^2$ --- see Table \ref{table:params}(a) --- show that
the fits are extremely bad, given the expected range, Eq.\
(\ref{eq:chi2.value}).

The nature of the bad fit is seen in the plots in the top row of each
of Figs.\ \ref{fig:Hippplots}--\ref{fig:PHCplots}.  Because the value
of $n_0$, the number of non-responding neurons, is much larger than
for the other $n_k$, we show two versions of the plot.  The right-hand
plots have the $k=0$ bin omitted and have a changed vertical scale, to
better exhibit the other bins.

The data considerably exceed the best fit in the $k=0$ bin in all 4
regions, but undershoot the fit in the other bins.  We have a choice.
If the sparsity is large enough to give a sufficient number of cells
that respond to multiple stimuli,
then, compared with data, too few cells are non-responsive.
If instead, the sparsity is low
enough to reproduce the number of non-responsive cells (i.e., $n_0$),
then there is much too small a probability for multiple responses.  In
either case, the model cannot reproduce all the data.

\subsection{One Active Population, One Silent Population Model}

A simple and natural improvement to the model is to add a population
of completely silent neurons that do not respond to any of the stimuli
used --- cf.\ \cite{dark.matter}.  That is, we make a 2-population
model with one active and one silent population of neurons.  The
silent population has sparsity zero.  For the active population, let
$\alpha_{\rm D}$ be its sparsity, and let $f_{\rm D}$ be its
fractional abundance. In this case, Eq.\ (\ref{random.neuron.multi})
becomes
\begin{equation}\label{twoparam}
\epsilon_k=f_{\rm D} \binom{S}{k} {\alpha_{\rm D}}^{k} (1-\alpha_{\rm D})^{S-k},
\end{equation}
for $k\geq1$. 

The results of a maximum-likelihood fit are shown numerically in Table
\ref{table:params}(b), and graphically in the middle row of each of Figs.\
\ref{fig:Hippplots}--\ref{fig:PHCplots}.  The fits are much improved,
but the $\chi^2$ values are still substantially too large for a good
fit.  Notice how the silent population contains by far the majority of
the neurons in all four regions.

The pattern of deviations between data and model is now an excess for
the data in the $k=1$ bins and a deficit at higher $k$.  That is, the
number of cells responding exactly once is substantially higher
compared with the extrapolation of the numbers of cells with multiple
responses.  This indicates that a better model would be to replace the
silent neural population by a slightly active population.  To fit the
data, this population must have a very small sparsity, so that it
predominantly gives contributions to the $k=0$ and $k=1$ bins only.

\subsection{Two-Population Model}

Therefore our final model uses two active populations each with a
particular sparsity.  One population we call the distributed
population, with a sparsity $\alpha_{\rm D}$ and fractional abundance
$f_{\rm D}$. 
The other population we call the ultra-sparse
population with sparsity $\alpha_{\rm US}$.  The fractional abundance of
the ultra-sparse population is $f_{\rm US}=1-f_{\rm D}$.  Then Eq.\
(\ref{random.neuron.multi}) becomes
\begin{multline} \label{threeparam}
\epsilon_{k} =
  (1-f_{\rm D}) \binom{S}{k} \alpha ^{k}_{\rm US} (1-\alpha_{\rm US}) ^{S-k} 
\\
  + f_{\rm D} \binom{S}{k} \alpha ^{k}_{\rm D} (1-\alpha_{\rm D}) ^{S-k}.
\end{multline}
The labeling of the populations is defined by $\alpha_{\rm US}<\alpha_{\rm D}$.  

The results of a maximum-likelihood fit are shown numerically in Table
\ref{table:params}(c), and graphically in the bottom row of each of
Figs.\ \ref{fig:Hippplots}--\ref{fig:PHCplots}.  The fits are much
improved.  For both the hippocampus and the entorhinal cortex, we have
good fits, with the model being consistent with the data.  The fit in
the amygdala is less reliable and the model poorly fits the data in
the parahippocampal cortex.  In all cases, the ultra-sparse population
is in the vast majority, around $90\%$ or more, while at the same time
having a very small sparsity, $10^{-3}$ or smaller.  Thus each neuron
in the ultra-sparse population responds on average to at most about
0.1 stimuli in a session.  We only see the effects of the ultra-sparse
population because the data are from a large number of neurons.  In
contrast, the remaining few percent of neurons in the other population
typically respond to several stimuli in each session.

For the parahippocampal cortex (PHC), a different or more general
model is clearly needed.  We observe that the functionality of the PHC
is much different than that of the hippocampus and the entorhinal
cortex, so it is not surprising that its neural coding properties
should be different.  The hippocampus is the classical locus of
episodic memory storage, and the entorhinal cortex is its main source
of input (and output).

\begin{figure*}
\centering
\begin{tabular}{lr}
  \includegraphics[scale=0.3]{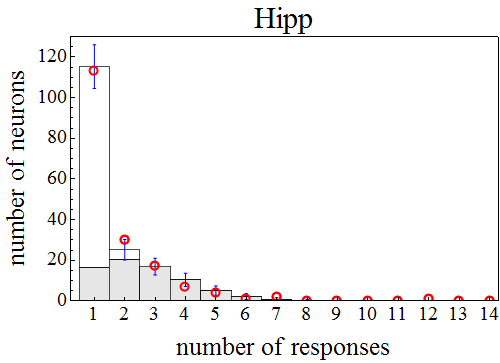} &  \includegraphics[scale=0.3]{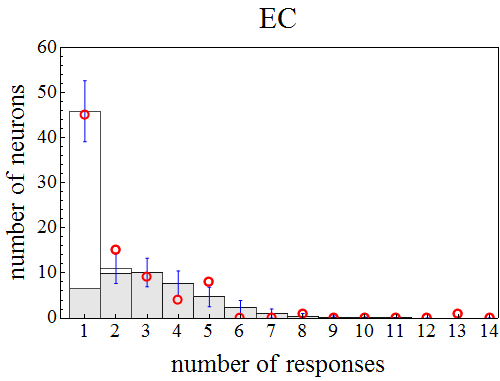} \\
  \includegraphics[scale=0.3]{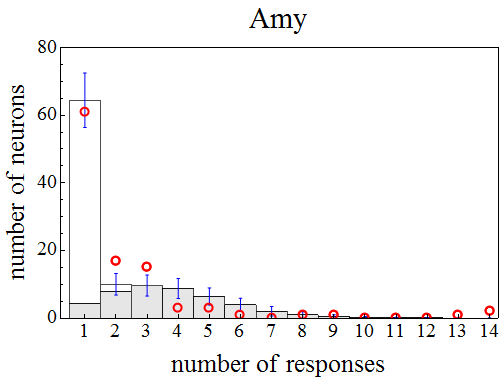}  & \includegraphics[scale=0.3]{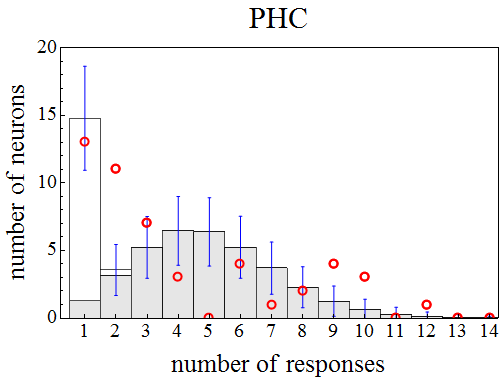} 
\end{tabular}
\caption{Plots of neuron responses predicted by the three-parameter
  maximum-likelihood fits to data in four regions of the MTL. The
  plots are of number of neurons as a function of number of
  responses. The shaded bars represent neurons in the almost-silent
  population while the open bars correspond to the distributed
  population. The red circles indicate the experimental values.} 
\label{fig:3param.stack}
\end{figure*}

An alternative view of the fit is shown in Fig.\
\ref{fig:3param.stack}.  Here we show how the neural responses are
predicted by the model to arise from the different populations.  The
bottom parts of the bars, shaded gray, show the expectation values for
the part of $N_k$ coming from above-threshold responses by neurons in
the D population.  Stacked above these are open bars, showing the
contribution from the US population.  In the bins with more than one
response, i.e., $k\geq2$, almost all the responses are from the D
population, with only a small contamination from the ultra-sparse
population, primarily at $k=2$.  In contrast, in the $k=1$ bin, there
is a relatively small fraction of responses from the D population,
from the tail of a distribution with its peak at several responses.
The majority of the $k=1$ bin is from the ultra-sparse population.
However, this represents only the tip of the iceberg, so to speak.
The vast majority of the ultra-sparse neurons give no above-threshold
responses; they appear in the $k=0$ bin, which is much too tall to be
shown in Fig.\ \ref{fig:3param.stack}.  

The sparsity and fraction for the D population can be determined from
the $k\geq2$ bins, i.e., with the exclusion of the $k=1$ bin.  There
are several bins involved, so the shape of the distribution of $N_k$
from a single sparsity fit is confirmed, as can be seen from the
lowest plots in the right-hand column of Figs.\
\ref{fig:Hippplots}--\ref{fig:PHCplots}, certainly for the HC and EC
regions. Extrapolating the fit for the D population to the $k=1$ falls
far short of the data, by a factor of at least 5.  This then
determines that there is an ultra-sparse population, whose average
sparsity is determined to a good approximation by the excess in the
$k=1$ bin relative to the total number of non-D neurons:
\begin{equation}
  \alpha_{\rm US}
  \simeq 
  \frac{ N_1(\text{excess above extrapolation}) }
       {  N(1-f_D) S}.
\end{equation}
Our actual best fit allows for the contamination of the bins of higher
$k$ by the ultra-sparse population.  The existence and size of the
ultra-sparse population is determined by the large excess of the
measured value of $N_1$ compared with the extrapolation from the bins
of larger $k$, whose relative sizes correspond to a sparsity of a few
per cent.

\subsection{Uncertainties and correlations in fitted parameters}

We computed uncertainties and correlations in the fitted values of the
models' parameters by the method described in Sec.\
\ref{sec:error.analysis}. 
 The uncertainties are reported in Table
\ref{table:params}.
Correlations between the three parameters of the full two-population
model were calculated using Eq.\ (\ref{correlations}), and are shown in
Table \ref{table:correlations}.

\begin{table}
\begin{tabular}{c|c|c|c}
		& $\rho_{\alpha_{US},\alpha_{D}}$ & $\rho_{\alpha_{US},f_D}$ & $\rho_{\alpha_{D},f_D}$ \\ \hline
	Hipp& 0.46 & -0.52 & -0.62 \\ \hline
	EC  & 0.34 & -0.33 & -0.41 \\ \hline
	Amy & 0.33 & -0.31 & -0.37 \\ \hline
	PHC & 0.30 & -0.23 & -0.19 \\ \hline
	\end{tabular}
\caption{Correlations between parameters of the two-active population
  model, assuming all recorded units are comprised of a single
  neuron. The presence of units consisting of two neurons did not
  affect the correlations between parameters to within two significant
  digits.} 
\label{table:correlations}
\end{table}

\subsection{Comparisons of the three versions of the model}

We can see that the results of fitting the first single-population
model with its single sparsity were intermediate sparsities
compromising between the extremes of the two populations in the full
model.  The single-population model therefore incorrectly represents
the actual neural sparsity.  Our fitted values of sparsity in Table
\ref{table:params}(a) roughly match those found by Waydo et al.\ \cite{Waydo}
in their fit of a pure one-population model to similar data.

In the second model, with a set of exactly silent neurons, the fit for
the responsive neurons is qualitatively similar to the D neurons in
the full model: a sparsity of a percent to a few percent and
a minority abundance.  Relative to the full model, the
value for the active population's sparsity is still biased downwards,
while its fractional abundance is biased upwards; these properties
give a compromise between the effects of the two populations of
neurons in the full model.

Merely introducing extra parameters increases the goodness of fit, but
only by an expected decrease of one unit in $\chi^2$ per parameter, as
in Eq.\ (\ref{eq:chi2.value}).  So the improved fits from adding an
extra population are highly significant.  In all cases, the $p$-values
for the poor fits for the first two models are well below $0.001$,
from standard plots or tables for the $\chi^2$ distribution.

\section{Multi-neuron Unit Results}
\label{sec:multi}

The results presented thus far have been under the assumption that all
recorded units are composed of a single neuron. However, it is known
that some units are in fact composed of multiple neurons \cite{GMC,
  Mormann}.  To gain an idea of the effect of such multiunits on our
fits and of our conclusions about neural properties, we apply the
general analysis from Sec.\ \ref{sec:multi.basics}.

\subsection{Implementation of multi-unit model}

To fit the data taking into consideration the presence of multi-neuron
units, we must maximize the likelihood function, Eq.\
(\ref{likelihoodfinal}) after replacing the $\epsilon_k$, with the
function $\epsilon^{\rm unit}_k$, Eq.\ (\ref{eq.multiunit.dk}). In the
case of the model with two populations at the neural level, i.e., with
$M=2$, Eq.\ (\ref{eq.multiunit.dkR})
becomes 
\begin{equation}
    \label{eq.multiunit.m2}
    \epsilon^{\rm unit}_{k,R} 
    = \sum_{l=0}^R \binom{R}{l} f_{\rm D}^l f_{\rm US}^{R-l}
      \, \binom{S}{k} (\alpha'_{l,R})^k (1-\alpha'_{l,R})^{S-k},
\end{equation}
where
\begin{equation}
    \label{eq.multiunit.alpha}
    \alpha'_{l,R}
    = 1-\left(1-\alpha_{\rm D}\right)^l\left(1-\alpha_{\rm US}\right)^{R-l},
\end{equation}
and, of course $f_{\rm US}=1-f_{\rm D}$.  Here $l$ and $R-l$ are the
numbers of neurons in the unit that are in the D and US populations. 
The result is that at the
unit level, we have multiple populations, each with its distinct
sparsity.  The different populations correspond to the different terms
in the summation in Eq.\ (\ref{eq.multiunit.m2}).  

For units with $R=1$, these just correspond to the original two neural
populations. For $R=2$, there are three populations.  One of these,
the most common case, is where both neurons in the unit are US
neurons, giving a sparsity $1-(1-\alpha_{\rm US})^2 \simeq
2\alpha_{\rm US}$, twice that of a single neuron.  The second most
common situation is where one neuron is in each neural population;
these units have sparsity $1-\left(1-\alpha_{\rm
    D}\right)\left(1-\alpha_{\rm US}\right) \simeq \alpha_{\rm D} +
\alpha_{\rm US}\simeq \alpha_{\rm D}$, where the last approximation
follows from the fact, confirmed by our detailed fit later, that
$\alpha_{\rm US}$ is much less than $\alpha_{\rm D}$.  The third
population, a small fraction $f_D^2$ of the units, has a larger
sparsity $1-(1-\alpha_{\rm D})^2 \simeq 2\alpha_{\rm D}$.

If only single-neuron and double-neuron units existed, i.e., if only
$R=1$ and $R=2$ occur, then the total number of populations at the
unit level would be 5, and these would appear in the formula
(\ref{eq.multiunit.dk}) for $\epsilon^{\rm unit}_k$.  If larger
multi-units occur, 
there are even more populations of units, each with its particular
sparsity.  This seems like a very complicated situation, but provides
no issue of principle in the MLE of the parameters of the populations
at the neural level, except that there is little data about the exact
distribution of the number of neurons in a unit.

However, as we will see in more detail shortly, considerable
simplifications occur because the vast majority of neurons are
extremely sparsely firing.  This property is reflected at the unit
level, and we will see that from a 2-population model at the neural
level, the unit-level data are reasonably accurately given by a
2-population model at the unit level.  This justifies a posteriori the
success of a 2-population model applied to unit level data, and one
can see how to relate properties of the neural populations to
properties of the unit populations.  The reasons come from the two
most common kind of unit.  The most common situation is that all the
neurons in a unit are all ultra-sparse, so that the unit itself
responds ultra-sparsely, typically at most one neuron at a time.  The
second most common situation is where exactly one of the neurons in a
unit is in the D population.  Then by far the most common response
from the unit is due to the single D neuron.

\subsection{Fit with multi-units}

\begin{table*}
  \begin{tabular}{c|c|c|c|c}
                           & Hipp                           &  EC                          & Amy                           & PHC                         \\ \hline
    $\alpha_{\rm D}$        & $(2.4\pm0.3)\times10^{-2}$     & $(3.0\pm0.4)\times10^{-2}$    & $(3.4\pm0.4)\times10^{-2}$  & $(4.7\pm0.5)\times10^{-2}$    \\ \hline
    $f_{\rm D}$             & $0.04\pm0.008$                 & $0.03\pm0.006$               & $0.03\pm0.006$                & $0.08\pm0.014$              \\ \hline
    $\alpha_{\rm US}$       & $(6.0\pm0.8)\times10^{-4}$      & $(3.2\pm0.6)\times10^{-4}$    & $(4.2\pm0.7)\times10^{-4}$  & $(3.3\pm1.2)\times10^{-4}$    \\ \hline
    $f_{\rm US}$           & $0.96\pm0.008$                 & $0.97\pm0.006$               & $0.97\pm0.006$                & $0.92\pm0.014$              \\ \hline
    $\chi^2(5)$            & $1.5$                          & $3.0$                        & $7.5$                         & $14$                        \\ \hline
    $\chi^2(10)$           & $4.9$                          & $10.1$                       & $15$                          & $31$                        \\ \hline
  \end{tabular}
  \caption{MLE values and $\chi^2$ values for two-population model, with $0.34N$ single units in each region and $0.66N$ double units in each region.}
\label{table:multiunit}
\end{table*}

\begin{figure*}
\centering
\begin{tabular}{cc}
    \includegraphics[scale=0.4]{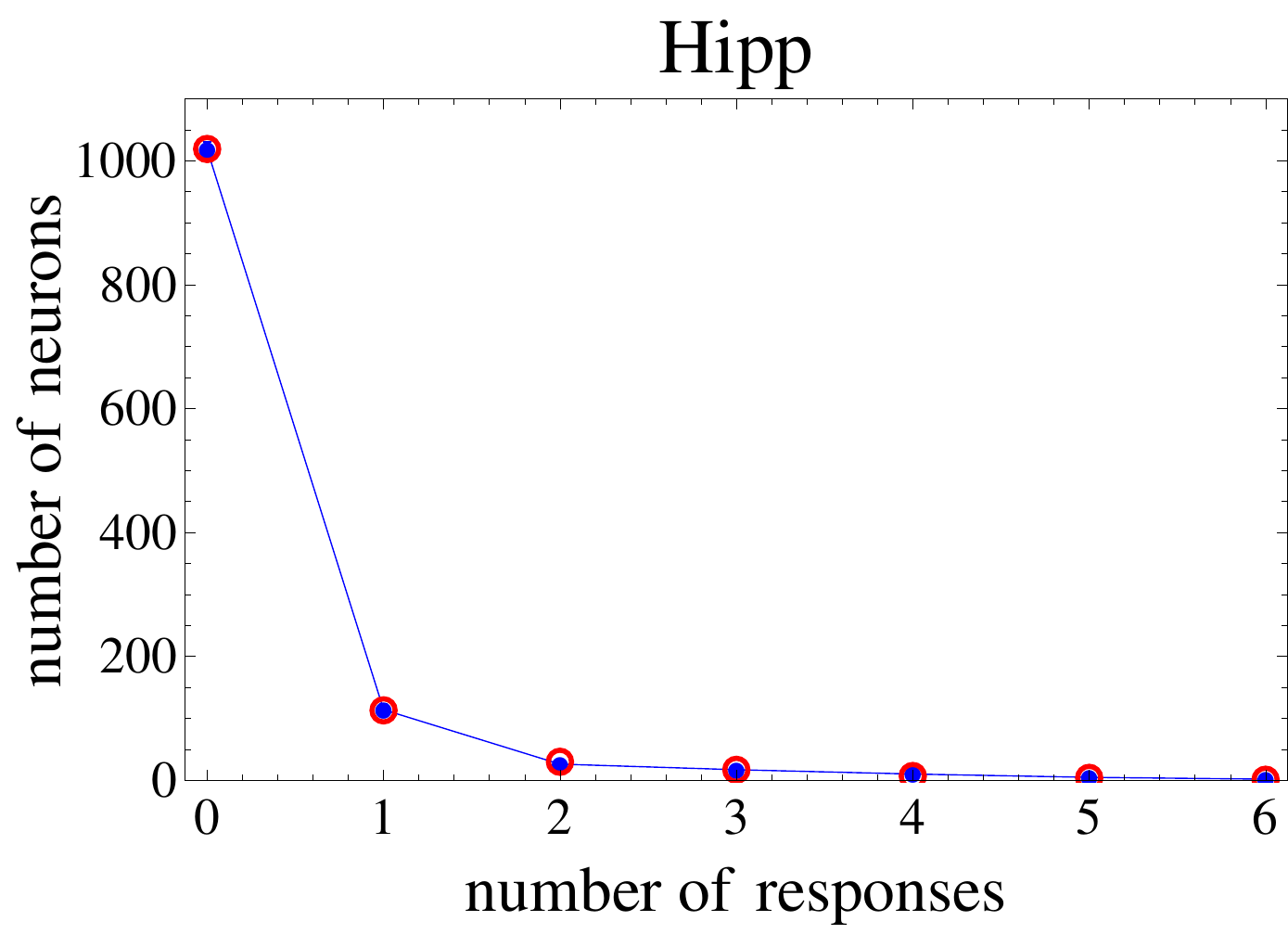} &  \includegraphics[scale=0.4]{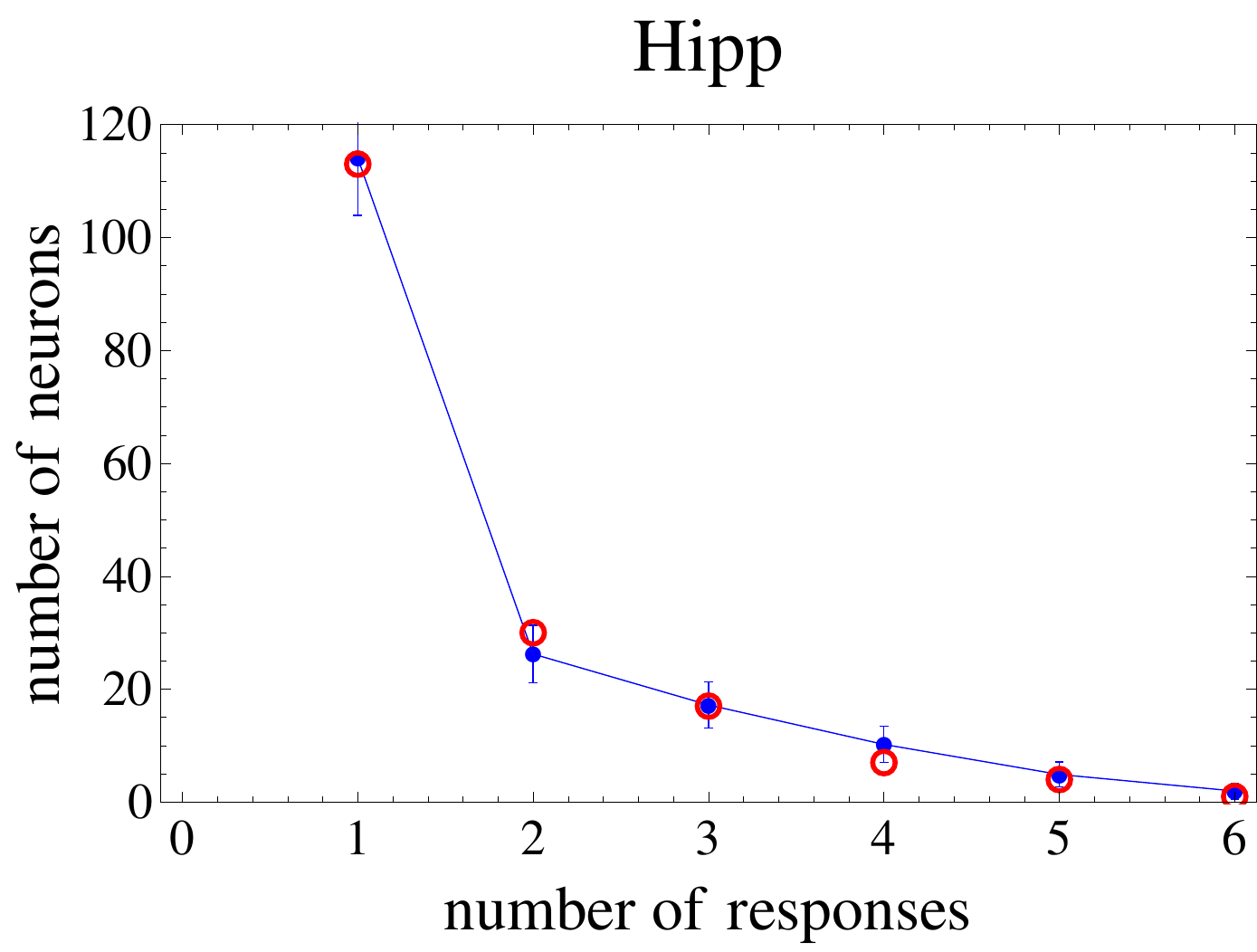}  
\end{tabular}
\caption{Plots of MLE fits to the Hipp with the multi-unit model.}
\label{fig.multiunit}
\end{figure*}

To understand how this works, we make a simplified model in which we
assume that $R\leq2$, i.e., that all measured units are either
composed of a single neuron or of two neurons. Let $p$ be the fraction
of single units. Then, we have for the total probability of a unit
responding to $k$ stimuli, $\epsilon^{\rm unit}_k$
\begin{equation}
	\label{eq.multiunit.m2t}
	\epsilon^{\rm unit}_k=p\epsilon^{\rm unit}_{k,1}+(1-p)\epsilon^{\rm unit}_{k,2}.
\end{equation}
From the number of single units reported in \cite{GMC}, we estimate
$p=0.34$, and we will use this value from now on.

We then used this in the formalism that we have already set up.  The
resulting fitted values of the parameters and the $\chi^2$ values are
shown in Table \ref{table:multiunit} and, for the case of the
hippocampus, the plots of the predicted and recorded $n_k$ are shown
in Fig.\ \ref{fig.multiunit}. Compared to the values fitted for the
original two-population model, Table \ref{table:params}(c), the
$\chi^2$ did not change greatly, although the goodness of fit is
somewhat improved, notably in the amygdala and PHC.

Thus the multi-unit model fits all four regions at least as well as
the basic two-population model, without any extra fitted parameters.
However, the values of some of the parameters did change.  Most
notably, the estimate of $\alpha_{\rm US}$ decreased by about
40\%. The earlier value is simply a weighted average of the effect of
single units containing one US neuron with sparsity $\alpha_{\rm US}$
and double units containing two US neurons, with unit sparsity
$2\alpha_{\rm US}$.  The fitting of the US population is determined
primarily from the $n_0$ and $n_1$ bins, so the response data by
itself cannot significantly determine the existence of these two
populations of units; it just gives a weighted average of the
sparsities.

In contrast, the value of $\alpha_{\rm D}$ is only slightly lower than
in the earlier fit; this is because most of the relevant data concerns
units that have one D neuron.  Given the sparseness at the unit level,
as in Table \ref{table:params}(c), which is tied to measured data, the
neural sparsity must be less when there are multiunits.

As to the population fractions, the value $f_{\rm D}$ is reduced by a
factor of about two thirds.  This is because for a given $f_{\rm D}$
at the neural level, there are two chances of having a D neuron in a
double unit.  Thus the effective value of $f_{\rm D}$ at the unit
level is the following weighted average:
\begin{equation}
  f_{\rm D} (1-p)  + 2 f_{\rm D} p
  = f_{\rm D} (1+p),
\end{equation}
which determines the difference between the values of $f_{\rm D}$ in
Tables \ref{table:params}(c) and \ref{table:multiunit} fairly well,
given the value $p=0.34$ that we deduced from Ref.\ \cite{Mormann}.

\subsection{Overall view of effects of multi-units}

We see that allowing for multi-units has actually strengthened our
conclusion that there is a large fraction of extremely sparsely
responding neurons.  Effectively the existence of multi-units has
diluted the effect in the data relative to the situation at the neural
level.  The original 2-population fit in Table \ref{table:params}(c)
characterizes the measured data at the unit level.  At the neural
level, as supported by Table \ref{table:multiunit}, the fraction of US
neurons must be even closer to unity, and their sparsity substantially
less than the already small values at the unit level.  This
qualitative result is independent of the exact numbers of multi-units
and the distribution of numbers of neurons in a unit.

\section{Discussion}
\label{sec:discuss}

Our primary result is that in the hippocampus\footnote{In the paper
  reporting the data we use, it is not stated which hippocampal region
  the recorded cells belong to.} (and other areas of the MTL), a vast
majority (90\% or higher) of neurons respond ultra-sparsely to stimuli
in the class presented --- around one in a thousand stimuli, or even
less.  Those neurons that respond more readily, i.e., to a few per
cent of the stimuli, comprise a rather small fraction (several per
cent) of the cells.

We have devised methods that treat the measured neurons and stimuli as
statistical samples, and allow the deduction of properties of the
ultra-sparse population even though these neurons respond on average
to less than a tenth of a stimulus out of the approximately 100 used
to obtain the data analyzed \cite{Mormann}.

There is the widely-known issue \cite{dark.matter, Henze,
  silent.cells, Olshausen.Field, HVC-RA} that in many areas of the
brain, there appear to be many silent or almost silent cells, i.e.,
cells that gave no detectable spikes at all in particular experimental
situations or that spike very rarely.  This gives the problem
\cite{Olshausen.Field} that responsive neurons reported in the
literature can be a very biased sample of all neurons in a probed
brain region. Our methods (and potential future improvements) give a
way of quantitatively measuring and correcting the bias.

\subsection{Choice of above-threshold responsiveness to define sparsity}

Care is needed in interpreting our results.  Most importantly, the
choice to classify a neuron binarily, as responsive or not, is at its
most natural when the neuron normally has a low firing rate and has
substantially higher activity under particular circumstances, with
fairly few border-line cases.

The simplest case is, of course, when the neuron is strictly binary.
That is, it gives exactly zero spikes under normal circumstances, and
gives several spikes only in one situation, as for the
$\mbox{HVC}_{\rm RA}$ neurons for zebra finches in \cite{HVC-RA}.  But
the binary classification is also sensible for cells such as the
pyramidal neuron in a human hippocampus whose responses are shown in
Fig.\ 3A of \cite{ison}.  It gives a consistently high response only
to a soaring eagle picture, out of those presented; nevertheless, its
response to other presented pictures, while small, is non-zero.

We do not need to commit to the exact semantics of such a neuron to
say that the categorical information coded in the high response of the
neuron is useful to the subject.  This information can be readily used
for read-out \cite{Buzsaki.2010} by a downstream neuron, and therefore
used to guide further behavior, etc.

In contrast, for an interneuron such as the one in Fig.\ 3B of
\cite{ison}, the application of a response threshold is much less
sensible.  For this neuron the normal state is a fairly high firing
rate somewhat correlated with the stimulus and its presentation.  One
can imagine occasional excursions above some chosen threshold, e.g.,
the 5-$\sigma$ threshold used in \cite{Mormann}.  But if these are only
small excursions, their meaning and utility is not obvious.  Of
course, if under some other specific circumstances not probed in the
reported data, the interneuron had a much higher firing rate, then it
would be natural to use a thresholded response criterion.  This is a
situation which our statistical methods are designed to address; if
the cell is typical of a certain class of cells, then one would see
examples of the rare large excursions of firing rate in other
cell-stimulus combinations.

The above remarks imply that for a better application of our methods,
one should 
extend the criteria for the responsive cells and their responses.  Not
merely should there be a number of spikes above some threshold
relative to the pre-stimulus situation, but the above-threshold
distribution of spike numbers should go well above the threshold.  It
would also be useful to classify cells by their firing rate in the
non-responsive state, so that our analysis by populations and sparsity
is applied to cells that are similar in characteristics.

\subsection{Populations}

We have identified two very different populations of cells in the
areas examined.  It is possible that these populations are
anatomically or functionally different, as with the different kinds of
cells in the HVC and RA areas of zebra finches \cite{HVC-RA,
  HVC-RAX}. But this is not a necessary deduction.  Indeed, we think
that is unlikely that our two populations correspond directly to the
pyramidal cell and interneuron populations reported in \cite{ison}.
The presence of the different populations might simply represent
different semantic properties for 
the cells' firing relative to the nature of the particular class of
stimuli used.  One can at best assume only that the situation is
typical.  For example, suppose one chose stimuli in a different class,
e.g., musical tunes instead of famous people.  Then some cells that
had very low sparsity in the famous-people situation could have much
higher sparsity in the music situation, and vice versa.  Since such
changing contexts are common experience, it is reasonable that this
situation is typical.  The different populations correspond to
different semantic domains, and the chosen stimuli sample these
domains.

The populations and their sparsities must then be treated as being
relative to the general class of stimuli used, e.g., famous
individuals, landmarks, animals or objects in the case of the data
from \cite{Mormann} that we analyzed.

\subsection{Cell semantics}
\label{sec:represention}

It has been suggested that the neurons under discussion are concept
cells \cite{concept.cells}; each cell responds to the presence, in
some sense, of a particular concept in the current stimulus.  One part
of the motivation for this is that the responses often appear to be
genuinely conceptual; for example, a cell might respond (within the
experimental data) only to stimuli involving a particular person,
e.g., to multiple different pictures of the person, even in disguise,
to the written name of the person, and to the spoken name.

In this section, we make some suggestions about how our results
quantitatively impact this issue.  We label the subject that of the
semantics of the cells, i.e., of what meaning should be attached to
their responses.

In the first place our results strengthen the basic case for
conceptual cells, by showing how sparse the responses typically are.
However, this case can only be made in conjunction with the
measurements of the conceptual properties of the responses.  To
understand this more clearly, it is important to recall that there are
two kinds of measurement involved.  

First, there are screening sessions, with many different unrelated
stimuli. The number of different objects and people averaged to 97,
and it was our analysis of this data that led to our deduction that
there are many very sparsely firing cells.

The screening sessions find pairs of stimulus and neuron where a high
response is obtained.  Then, in testing sessions, variants on the
response-causing stimuli were used.  It is the testing sessions that
establish that many responses are indeed conceptual --- e.g., a neuron
responds to a picture, the written name, and to the spoken name of a
particular human.  However, it is the screening session data that
determine the sparseness of the cells' responses, because the
screening sessions have the largest number of identifiably distinct
concepts.

It is tempting to say that because a cell consistently responds to a
variety of very different stimuli related to a particular person,
e.g., Jennifer Aniston \cite{GMC}, that the cell actual represents
Jennifer Aniston, i.e., that its firing above threshold codes that the
concept of Jennifer Aniston is being processed or has been detected.
This is the simplest version of the concept cell idea
\cite{concept.cells}.

We now address three questions about the exact correctness of this
interpretation: One is whether the concept involved is in fact the
obvious one, e.g., Jennifer Aniston in the case just mentioned.  The
second is whether the concept is one that in some sense corresponds to
the current stimulus, be it visual or auditory.  Our third question is
whether the neural conceptual representation is strictly local, i.e.,
whether the above-threshold firing of one of these cells codes only a
single concept.  Our discussion of these questions will provoke a
fourth question: Whether the representation for only one concept is
active at a given time is active or whether the representations of
multiple concepts are (more-or-less) simultaneously active, on a time
scale of a few hundred milliseconds. 

First it appears necessary that a concept, like Jennifer Aniston, is
coded and represented in the subject's brain.  But the reported cell
can be a downstream consequence.  For example, it might code the
activation of an episodic memory of a show in which she participated.
Evidently, such downstream activated concepts are related to the
Jennifer Aniston concept.  Support of the idea that the apparent
concept cells may code downstream concepts is that the Jennifer
Aniston cell was later found \cite{concept.cells} to respond to a
picture of Lisa Kudrow, a co-star of Aniston's in a television series.
That individual episodic memories may be among the concepts involved
(for some appropriate definition of ``concept'') is suggested because
of the well-known central role of the hippocampus in the episodic
memory system.

We know that, at least in humans, recall of memories can be based not
directly on a current stimulus but on cues generated by internal
cognition.  In general, this removes an obligatory link between
stimulus properties and conceptual neural responses. For the
experimental measurements under discussion, this issue need not be
important, because the experimental protocol was explicitly designed
to have the subjects' attention focused on the stimuli.

Once one allows that downstream concepts are activated, one should
expect that multiple concepts are simultaneously active, with perhaps
only one being selected at a given time for conscious attention.  This
is essentially the same property that computer search engines have.
We can regard these as associative memory systems.  A cue, e.g., a
search string, is provided as input, and the result is a list of items
containing or related to the cue; these are the activated concepts.
The user can click on one item in the list to get its content; this is
analogous to conscious memory recall. 

The next question is whether or not the representation is local.
Normally a dichotomy is made simply between local representations and
distributed representations.  In the case of distributed
representations, even when they are sparse, it is generally assumed
that the individual neurons that are involved in a distributed
representation of an object themselves represent features or
properties of the object in question --- see, for example, Ref.\
\cite{sparse.coding.Olshausen.Field,sparse.coding} and references
therein.  Such representations were called ``iconic'' by Wickelgren
\cite{Wickelgren.1969, Wickelgren.1992, Wickelgren.1999}.  

But a further possibility is the use of what Wickelgren
\cite{Wickelgren.1969, Wickelgren.1992, Wickelgren.1999} called
``chunk assemblies''.  Each of these is a relatively small set of
cells, the activation of which codes the presence or processing of the
associated concept.  The individual cells in a chunk assembly do not
themselves code features corresponding to the concept.  This is
provides an important modification to the concept-cell idea.
Quantitatively, coding using chunk assemblies is characterized by the
number of cells in each assembly and by the number of chunk assemblies
in which each cell participates (which need not be exactly fixed
numbers).  Local coding is the limiting case in which each cell
participates in the assembly for exactly one concept, as opposed to
participating in merely a very small fraction of the concepts stored
in a system.  We can regard our work as a step in determining
quantitative properties of chunk assemblies.  When only a small number
of stimuli are used, a cell in a chunk assembly will behave quite
similarly to a cell providing a local representation.

We now work out a relation between the sparsity of cell responses, and
some of the coding properties.  The properties of interest are the
total number of concepts coded, the number of concepts that each cell
codes (i.e., the number of chunk assemblies it participates in), and
the number of concepts that are simultaneously active.  Of course
neither of the last two numbers need be fixed, but it will be useful
to treat them as single or representative numbers to get an idea of
the relation to sparsity.  If only one concept were active at a time,
and if the coding were local (so there is only one concept per cell),
then the sparsity would be $1/\mbox{Total \# concepts}$.  

With the possibilities of multiple concepts being simultaneously
active and of more the one concept being coded per cell, we get
instead: 
\begin{equation}
\label{eq:sparsity.relation}
\mbox{sparsity} 
= \frac{ \left( \mbox{\# concepts activated} \right)
  \left( \mbox{\# concepts per cell} \right)
}
{ \mbox{Total \# concepts} },
\end{equation}
at least in some average sense, given that both the number of concepts activated and the number stored by the cell are small compared with total number of concepts stored overall.  The number of concepts per cell in
the case of a local representation is unity.  The simplest ideas about
local/grandmother-cell representations would also assign unity to the
number of simultaneously active concepts.  Given the number of
concepts we should expect to be represented in the human brain
(probably hundreds of thousands if not millions), even our small
measured sparsity is much too large to be consistent with a local
representation.  In this discussion, we should use the word
``concept'' very broadly, such that it includes, for example,
individual episodic memories.

Now an estimate of 10,000 to 30,000 has been quoted by Waydo et al.\
\cite{Waydo} as the number of objects that a typical adult recognizes,
on the basis of work by Biederman \cite{Biederman}.  But this is
surely a substantial underestimate of the number of concepts that are
coded in a human brain.  First, the number of words in a language that
known to an adult is in the tens of thousands, and the number of
concepts is surely substantially higher.  Second, it is known that
humans can remember thousands of pictures \cite{Standing,Konkle}
presented during a single day.  The total number of memories created
over a lifetime must be orders of magnitude larger.  Even allowing
that on a long time scale many of these will be forgotten, this
indicates that the number of concepts/objects represented can easily
be in the hundreds of thousands or millions.

For illustration assume that the number of concepts remembered is
$10^6$.  Then a measured sparsity of around $6\times10^{-4}$, as we found for
the majority population from our analysis, implies that
\begin{equation}
 \left( \mbox{\# concepts activated} \right)
 \times \left( \mbox{\# concepts per cell} \right)
 \simeq 600.
\end{equation}
We leave further analysis to the future, but use this estimate as a
suggestion about the quantitative properties of concept coding.

\subsection{Confirmation and relation to previous work}

Our results confirm and substantially strengthen results found by one
of us and Jin in \cite{CJ}.  There we used earlier data from
\cite{GMC} that only provided values for the number of units with one
response, $N_1$, the number with two or more responses, $N_{\geq2}$, and
for the average number of responses from responsive units.  We ruled
out a one-population model, and fit the three parameters of the
two-population model from the three reported summary statistics.  But
there was therefore no test of goodness of fit for the 3-parameter
model.  Instead the shape of the $N_k$ distribution was a prediction,
which is successfully tested in this paper, for the case the
hippocampus and entorhinal cortex, at least.

In the present paper,
we also give a more systematic account of the statistical methods, and
have a breakdown by brain region, using newer data.  Although the
exact parameters of the fits are a little different, the main picture
is confirmed and tested.  The differences could be accounted for by
differences in the subjects and of detailed experimental procedures
and by our different treatment of multiunits.

\subsection{Treves-Rolls definition of sparseness}

An alternative measure of sparseness proposed by Rolls and Tovee \cite{Rolls.Tovee.1995} and reviewed by Treves and Rolls \cite{Rolls.Treves.2011}, is calculated from the firing rates of neurons in response to stimuli. The average sparseness reported
\cite{Rolls.Treves.2011} with this definition was $0.34\pm0.13$ (for
hippocampal spatial view cells in the macaque hippocampus). This is
dramatically different than what we found with our definition of
sparsity.  It is worth understanding how this difference might arise
(aside from a conceivable difference between species).  We will 
demonstrate that the Treves-Rolls definition can be very misleading
as to the nature of neural coding.

They define the sparseness of a neuron in response to a sample of $S$
stimuli as
\begin{equation}
  \label{eq:R.T.sparseness}
  a_{\rm TR}^s = \frac{ \langle r \rangle^2 }{ \langle r^2 \rangle }
      = \frac{ \left( \sum_{j=1}r_j/S \right)^2  }
             { \left( \sum_{j=1}r_j^2 \right) / S  },
\end{equation}
where $r_j$ is the mean firing rate of the neuron in response to
stimulus $j$ and $\langle \ldots \rangle$ denotes an average of a quantity over all
presented stimuli. 
In the case that the neuron is binary in its responses, i.e., that it gives
a large response with some fixed firing rate to some stimuli and is
exactly silent for all other stimuli, this definition gives the same
result as our definition of sparsity.  

Note, however, that there is a difference that the above definition is
applied to one neuron with the stimuli actually presented in an
experiment, whereas ours applies for a universe of stimuli.  Our
sparsity is something that must be inferred from statistical arguments
applied to data, whereas theirs is directly defined as a property of
the data.  Even so, if neurons are \emph{exactly} binary and all have
the same sparsity (in our sense), then the Treves-Rolls sparsity is a
good estimator of our sparsity.

The definition (\ref{eq:R.T.sparseness}) is in fact the unique
combination of the first and second moments of the firing rate that
gives $\alpha$ for exactly binary neurons (always in the limit of large
$S$).  However, as we will now show, Rolls-Treves sparseness and our
thresholded sparsity can have widely different values if the neurons
are not strictly binary.

We first observe that Rolls and Treves only reported a value of
sparsity averaged over all neurons.  Now Ison et al.\ \cite{ison} in
their Fig.\ S2 showed the distribution across neurons of various
measures of sparseness and selectivity.  There is a wide range of
Rolls-Treves sparseness from less than 0.1 (the most common) to
another peak close to 1 (for putative interneurons).  The average of
this distribution is quite misleading as to the properties of
individual neurons.  Furthermore, all the distributions in that figure
are measures of sparseness with respect to the presented stimuli, and
no attempt is made to infer an underlying sparseness or selectivity
defined with respect to a whole class of stimuli, such as we do here
and Waydo et al did in \cite{Waydo}.

It is well known that many (but not all) hippocampal neurons respond
strongly to certain specific behaviorally relevant stimuli or
situations, while responding weakly or not at all at other times.  The
data we analyzed simply give a particularly notable example.  Let us
refer to the situations to which a cell responds strongly as
``on-target'' and the other situations as ``off-target''.  Suppose,
that the on-target responses are very rare, as we have found, but that
the off-target responses, while being of low rate, are nevertheless
non-zero. For example, the hippocampal place cells found in rats are
known to have dramatically different firing rates when an animal is
in the place field of the cell (producing an on-target response)
compared to when the subject is out of the place field (producing an
off-target response).

We will now point out by constructing an appropriate class of models
that the Treves-Rolls
sparseness can be dominated by properties of the off-target firing while
being quite insensitive to the on-target responses.  At the same time,
an appropriate choice of threshold can make it quite unambiguous as
to which responses are on-target and which are off-target. The model
is intended not to be (at this point) an actual
representation of data, but just a reasonable counterexample, to show
how a high average value of Rolls-Treves sparseness can be compatible
with a very low sparsity in our sense.

In this model each neuron has a categorical response to a certain
class of stimuli, with probability $\alpha$.  But instead of assuming
purely binary neurons, we postulate a pseudo-binary model in which the on-target and off-target
firing both come from a distribution of firing rates, but with
different distributions.  Let us assume that the off-target and
on-target firing rates have the distributions $P_0(r)$ and $P_1(r)$
respectively.\footnote{Analysis of a large, unbiased set of neurons in the rat MTL reported in \cite{mizuseki.buzsaki} suggests a log-normal distribution for both $P_0(r)$ and $P_1(r)$. However, the exact forms of the distributions $P_0(r)$ and $P_1(r)$ are not important in this analysis, so long as they are largely non-overlapping.} Then the total distribution is a mixture:
\begin{equation}
  \label{eq:P.r}
  P(r) = (1-\alpha) P_0(r) + \alpha P_1(r).
\end{equation}
We let $r_0$ and $\Delta r_0$ be the mean and standard deviation of
the off-target responses, and let $r_1$ and $\Delta r_1$ be the same
quantities for the on-target responses. Naturally, we should assume
that $r_1$ is sufficiently much higher than $r_0$ that on-target
responses can be detected adequately reliably; this depends on the
tail of $P_0(r)$ falling sufficiently rapidly as $r$ increases.

A calculation yields the Rolls-Treves sparseness (relative to all
stimuli) as:
\begin{multline}
  \label{eq:a.TR.calc}
  a_{\rm TR}^s
\\
  = \frac{ \displaystyle
           \left( 1
                   + \frac{\alpha}{1-\alpha} \frac{r_1}{r_0}
           \right)^2
         }
         { \displaystyle
           \frac{1}{1-\alpha} 
             \left( 1 + \frac{(\Delta r_0)^2}{r_0^2} \right)
           + \frac{\alpha}{ (1-\alpha)^2 }
             \left( \frac{r_1^2}{r_0^2} 
                  + \frac{(\Delta r_1)^2}{r_0^2} 
             \right)
         }.
\end{multline}
This reproduces the value $\alpha$ in the case of binary neurons, i.e.,
where the standard deviations are negligible and the limit $r_0 \to 0$
is taken.  But if instead we take the situation that $\alpha$ is very small
(much less than the ratio $r_0/r_1$ of off- to on-target mean
responses), then the Treves-Rolls sparseness approaches $1/(1+ \Delta
r_0^2/r_0^2)$.  This is just the Rolls-Treves sparseness calculated
purely from the off-target distribution.  In this case the
Rolls-Treves sparseness says nothing about the on-target responses.

If the two distributions $P_0$ and $P_1$ are distinct enough, then it
is possible to set a threshold $r_{\rm th}$ to give reliable detection
of on-target and off-target states.  Given an off-target stimulus, we
need the false positive rate to be substantially less than $\alpha$, so
that above threshold responses are predominantly for on-target
stimuli.  This requires
\begin{equation}
  \int_{r_{\rm th}}^\infty P_0(r) dr  \ll \alpha,
\end{equation}
i.e., that the threshold is sufficiently far out on the tail of $P_0$.

Given an on-target stimulus, we need it to be reliably detected, so
that the false negative rate (relative to $\alpha$) is small, i.e.,
\begin{equation}
  \int_0^{r_{\rm th}} P_1(r) dr  \ll 1.
\end{equation}
This is arranged if the bulk of the on-target distribution is beyond
the threshold.

For any given value of $\alpha$ we can arrange to satisfy all these
conditions with appropriate functions for the two underlying
distributions $P_0(r)$ and $P_1(r)$.  That is, the Rolls-Treves
sparseness can be dominated by its off-target value,
while we have reliable discrimination between on-target and off-target
stimuli, despite the low proportion $\alpha$ of on-target stimuli.

In this situation, there is no contradiction between a rather high
value of Rolls-Treves sparseness, like $0.3$ and extremely low values
of our sparsity.  Given the behavioral significance of the
above-threshold responses in \cite{GMC}, it is sparsity in our sense
that is most relevant to understanding the nature of the corresponding
conceptual coding.

\subsection{What properties of neural responses are determined?}

Although our model provides a good fit to the data (at least in the
hippocampus and entorhinal cortex), it should not be supposed that the
model gives an exact characterization of neural responses to stimuli.

The first issue is simply that if we replaced one particular neural
population of a particular sparsity $\alpha$ and fractional abundance
$f$ by several populations with sparsities not too far from the
original single value, and with a total abundance summing to $f$, the
result would not be very distinguishable from the original case.  In
the limit of a large number of stimuli, the number of responses
from one population
would be clustered at $k=\alpha S$, with a standard deviation
$\sqrt{\alpha S}$ that is much smaller than the mean.  The
\emph{fractional} standard deviation is $1/\sqrt{\alpha S}$.  But with
the actual values of $S$ and $\alpha$, the standard deviation is
comparable to the mean; indeed, for the ultra-sparse population the
standard deviation is much larger than the mean.  Thus splitting one
population into a set of populations nearby in sparsity produces
little measurable effect.

The lack of ability to distinguish populations of nearby sparsity is
particularly notable for the ultra-sparse population.  This populates
primarily the $k=1$ bin.  Thus our measurement of a sparsity for the
ultra-sparse population is really a measurement of a weighted average
of the sparsities of the ultra-sparsely responding neurons.  

What is properly deduced from our analysis is that there are
relatively few neurons that respond with a sparsity of a few per cent,
and a much larger number that respond much more sparsely.

A more precise understanding of how the multiunit cases arise from an
underlying neural response --- e.g., \cite{extracell} --- would lead
to a more precise estimation of the population parameters at the
neural level.

A second issue \cite{dark.matter, Henze, silent.cells,
  Olshausen.Field, HVC-RA} is that there may be many silent cells,
i.e., cells that gave no detectable spikes at all in the measurements.
Silent cells are to be contrasted with the many non-responsive cells
that were included and were important in our analysis; non-responsive
cells did give detected spikes, but never enough to count as an
above-threshold response.  On the basis of results in \cite{Henze},
Waydo et al.\ \cite{Waydo} argue that this is potentially a very large
effect.  They suggest that ``as many as 120--140 neurons are within
the sampling region of a tetrode in the CA1 region of the
hippocampus'', but say that they only identified ``1--5 units per
electrode''.  Of course some of these units are multiunits,
corresponding to two or more neurons.

To quantify the effects of silent cells, let $K$ be the ratio the
total number of cells to the number of detected cells.  The suggestion
\cite{dark.matter, Henze, Waydo} is that $K$ could be as much as an order of
magnitude.  The effect of silent cells on our measurement of the
sparsity $\alpha_{\rm D}$ of the higher sparsity population would be
negligible.  This is simply because these cells typically respond to
multiple stimuli in the data, and are actually detected.  The value of
$\alpha_{\rm D}$ is obtained primarily from the relative numbers of
cells with $k=2$, $k=3$, etc responses.  But their fraction in the
total neuronal population, $f_{\rm D}$ must be decreased by a factor
$K$.

For the ultra-sparse population, the sparsity $\alpha_{\rm US}$ is
decreased relative to our fits by a factor $K$ (while the population
abundance gets closer to 100\%).  This is simply because $\alpha_{\rm
  US}$ is, to a first approximation, the ratio of $N_1$ to $N_0+N_1$
(after taking out the estimated contamination of these bins from the
other population).  $N_1$ is fixed by data, but $N_0+N_1$ is increased
by a factor of about $K$.  Effectively, if there is a population of
completely silent cells, then the measured number $N_0$ of cells that
gave no above-threshold responses should be changed to approximately
$KN_0$.  (The approximations consist of neglecting $N_1$ with respect
to $N_0$ and neglecting the small number of cells in the D population
for which all the stimuli were off-target.

The effect of possible large numbers of silent cells is to
substantially strengthen our conclusions, even if it makes some of the
numbers less certain.  Of course, if at sometime in the future, a more
precise understanding of electrode properties were obtained, then a
useful estimate of the silent cell factor $K$ could be obtained. After
that our numerical results could be adjusted accordingly.  See
\cite{spike.sorting.issue, extracell} for recent work on this issue.

A final issue is that the measurements of sparsity are relative to a
particular set of stimuli.  We need to regard the stimuli as being
chosen as a sample from an enormous number in a general subject area
known to the patient.  The general topics were chosen to correspond to
areas that were well known to the subjects.  Given that the neural
responses were very specific, one must suppose that if a very
different topic were chosen (e.g., scientific subjects instead of
movie stars), the responses (or lack thereof) by particular neurons
could be very different.  Some previously responsive cells might even
become silent, and vice versa.  Similar phenomena are seen with place
field behavior of hippocampal neurons when an animals environment is
changed \cite{lee.rao}.  So any estimate of the sparsity of the
response of a particular neuron is relative to the subject area of the
stimuli. Nevertheless, it is reasonable that the distribution of
sparsity across a population of hippocampal neurons would be not much
changed between different stimuli sets.

It is known that the hippocampus is involved generally in episodic
memory, so over the whole neural population one should expect to get
responses to stimuli of all subject areas.  There should not be
specialization for the particular subject areas used for the
measurements except for the fact that the topics are ones for which
the subjects have abundant memories.  Therefore our results concerning
the neuronal population as a whole should be regarded as broadly
applicable.  That is, we should expect similar results for other
topics for the stimuli.  This conclusion also makes it acceptable that
we analyzed data not just from one patient but data pooled over many
patients.

\appendix
\begin{widetext}
\section{Poisson approximation for likelihood}
\label{sec:Poisson}

We now review the derivation of the Poisson approximation
(\ref{likelihoodfinal}) to the likelihood function
(\ref{likelihood}). The derivation applies when the values of
$\epsilon_k$ for non-zero $k$ are small, more precisely, when
$\sum_{k=1}^S\epsilon_k \ll 1$.

We start from Eq.\ (\ref{likelihood}) by factoring out the $k=0$
factors from the product, and then using Eqs.\ (\ref{epnormal}) and
(\ref{Nsum}) to write $n_0$ and $\epsilon_0$ in terms of the
corresponding quantities for $k\geq 1$:
\begin{equation}\label{likelihood1}
  \mathcal{L}\xleft( \{ \alpha_i, f_i \} \right)
  = \frac{N!}{\left(N-n_{\geq1} \right)!}
     \left(1-\epsilon_{\geq1}\right)^{N-n_{\geq 1}}
     \prod_{k=1}^{S}{\frac{\epsilon_{k}^{n_k}}{n_k!}},
\end{equation}
where $n_{\geq 1}=\sum_{k=1}^{S}{n_k}$, and 
$\epsilon_{\geq 1}=\sum_{k=1}^{S}{\epsilon_k}$.
By taking the logarithm of both sides of Eq.\ (\ref{likelihood1}) we get:
\begin{equation}\label{log1}
  \ln\mathcal{L}
  = \ln{N!}
    - \ln[(N-n_{\geq1})!]
    + (N-n_{\geq1})\ln(1-\epsilon_{\geq1}) 
    + \ln\prod_{k=1}^{S}{\frac{\epsilon_{k}^{n_k}}{n_k!}}.
\end{equation}%
For large $N$, we can use Stirling's approximation to $O(1/N)$:
\begin{equation}
\ln{N!} = N\ln{N} - N + \frac{1}{2} \ln(2\pi N) + O\xleft( \frac{1}{N} \right).
\end{equation}%
Applying this formula to the first two terms in (\ref{log1}) yields 
\begin{align}
\label{loglongform}
\ln\mathcal{L} 
& =  N\ln{N} + (N-n_{\geq1})\ln\xleft(
     \frac{1-\epsilon_{\geq1}}{N-n_{\geq1}} \right) -  n_{\geq1} 
     +\ln\prod_{k=1}^{S}{\frac{\epsilon_{k}^{n_k}}{n_k!}} 
     + O\xleft( \frac{1}{N} \right)
\nonumber\\
&=  n_{\geq1}\ln{N} 
    + (N-n_{\geq1})
      \ln \xleft( 
               1 - \frac{\epsilon_{\geq1}-n_{\geq1}/N}{1-n_{\geq1}/N} 
          \right)
    - n_{\geq1}
    +\ln\prod_{k=1}^{S}{\frac{\epsilon_{k}^{n_k}}{n_k!}}
    + O\left( \frac{1}{N} \right).
\end{align}%
(The approximation worsens beyond the order $1/N$ error estimate if
$n_{\geq 1}$ gets close to $N$.  But since the $\epsilon_k$ are small,
this situation is of very low probability.  Henceforth the error
estimates will apply not too far from the peak of the likelihood
distribution.)

The first term can be combined with the product term
\begin{equation}\label{2terms}
  n_{\geq1}\ln{N} + \ln\prod_{k=1}^{S}{\frac{\epsilon_{k}^{n_k}}{n_k!}}
= \ln\left[{N^{n_{\geq1}}}\prod_{k=1}^{S}{\frac{\epsilon_{k}^{n_k}}{n_k!}}\right]
= \ln\prod_{k=1}^{S}{\frac{\left(N \epsilon_{k}\right)^{n_k}}{n_k!}}.
\end{equation}%
We can simplify the remaining logarithm in Eq.\ (\ref{loglongform})
\begin{equation}\label{logterm}
  (N-n_{\geq1})\ln{\left( 1-\frac{\epsilon_{\geq1}-n_{\geq1}/N}{1-n_{\geq1}/N} \right)} 
= \left(N-n_{\geq1}\right)
  \left(\frac{ n_{\geq1}/N - \epsilon_{\geq1} }{ 1-n_{\geq1}/N } 
     + O\xleft[\left(\frac{\epsilon_{\geq1}-n_{\geq1}/N}{1-n_{\geq1}/N}\right)^2\right]
  \right) .
\end{equation}%
For a multinomial distribution,
\begin{align}
  \epsilon_{\geq1}-n_{\geq1}/N
  = \sum_{k=1}^{S}{(\epsilon_k-n_k/N)}
  = \sum_{k=1}^{S} O\xleft(\sqrt{\epsilon_k/N}\right),
\end{align}%
where we have used the standard deviation of the distribution to
estimate the typical value of $|\epsilon_k-n_k/N|$.

Thus we can write:
\begin{equation}
  O\left[\left(\frac{\epsilon_{\geq1}-n_{\geq1}/N}{1-n_{\geq1}/N}\right)^2\right] 
= O\left[\frac{1}{N}\left(\sum_{k=1}^{S}{\sqrt{\epsilon_k}}\right)^2\right]
= O\left(\epsilon_{\geq1}/N\right).
\end{equation}%
Thus, Eq.\ (\ref{logterm}) can be simplified to:
\begin{equation}\label{logterm2}
  (N-n_{\geq1})\ln{\left( 1 - \frac{\epsilon_{\geq1}-n_{\geq1}/N}{1-n_{\geq1}/N} \right)}
=
  n_{\geq1}  - N\epsilon_{\geq1} + O(\epsilon_{\geq1}) + O(1/N)
\end{equation}
Substituting Eqs.\ (\ref{2terms}) and (\ref{logterm2}) into Eq.\ (\ref{loglongform}) yields
\begin{equation}
\ln\mathcal{L} 
= -N\epsilon_{\geq1} 
  + \ln\prod_{k=1}^{S}{\frac{\left(N \epsilon_{k}\right)^{n_k}}{n_k!}} 
  + O(\epsilon_{\geq1}) + O(1/N).
\end{equation}
Therefore the likelihood function in our approximation is given by a
product of Poisson distributions:
\begin{equation}\label{likelihoodfinal1}
\mathcal{L} \approx \prod_{k=1}^{S} {e^{-N \epsilon_{k}}{\frac{\left(N \epsilon_{k}\right)^{n_k}}{n_k!}}}.
\end{equation}

\end{widetext}




\begin{thebibliography}{99}

\bibitem[Attwell and Lauglin (2001)]{metabolism2}
   D. Attwell and S.B. Laughlin,
   ``An Energy Budget for Signaling in the Grey Matter of the Brain'',
   J. Cereb.\ Blood Flow Metab.\ \textbf{21}, 1133--1145 (2001).

\bibitem[Baker and Cousins (1984)]{Baker.Cousins}
   S. Baker and R.D. Cousins,
   ``Clarification of the Use of Chi-Square and Likelihood Functions
   in Fits to Histograms'',
   Nucl.\ Instrum.\ Meth.\ \textbf{221}, 437--442 (1984).

\bibitem[Biederman \ (1987)]{Biederman}
   I. Biederman,
   ``Recognition-by-Components: A Theory of Human Image Understanding'',
   Psychol.\ Rev.\ \textbf{94}, 115--147 (1987).

\bibitem[Bowers \ (2009)]{Bowers09}
  J. Bowers,
  ``On the Biological Plausibility of Grandmother Cells: Implications
  for Neural Network Theories in Psychology and Neuroscience'', 
  Psychol.\ Rev.\ \textbf{116}, 220--251 (2009).

\bibitem[Bowers \ (2010)]{Bowers10a}
  J. Bowers,
  ``More on Grandmother Cells and the Biological Implausibility of PDP
  Models of Cognition: A Reply to Plaut and McClelland (2010) and
  Quian Quiroga and Kreiman (2010)'', 
  Psychol.\ Rev.\ \textbf{117}, 300--306 (2010).

\bibitem[Bowers \ (2010)]{Bowers10b}
  J. Bowers,
  ``Postscript: Some Final Thoughts on Grandmother Cells, Distributed
  Representations, and PDP Models of Cognition'', 
  Psychol.\ Rev.\ \textbf{117}, 306--308 (2010).
  
\bibitem[Buzs\'aki, (2010)]{Buzsaki.2010}
  G. Buzs\'aki,
  ``Neural Syntax: Cell Assemblies, Synapsembles, and Readers'',
  Neuron \textbf{68}, 362--385 (2010).

\bibitem[Camu\~nas-Mesa and Quiroga (2013)]{extracell}
  L.A. Camu\~nas-Mesa and R.Q. Quiroga,
  ``A Detailed and Fast Model of Extracellular Recordings'',
  Neural Comput.\ \textbf{25}, 1191--1212 (2013).

\bibitem[Collins and Jin (2005)]{CJ}
   J. Collins and D.Z. Jin,
   ``Grandmother cells and the storage capacity of the human brain'',
  arXiv:q-bio/060301.

\bibitem[Foldiak and Endres (2008)]{sparse.coding}
   P. F\"oldi\'ak and D. Endres,
   ``Sparse coding'',
   Scholarpedia, 3(1):2984 (2008). 

\bibitem[Hahnloser, Kozhevnikov and Fee (2002)]{HVC-RA}
   R.H.R. Hahnloser, A.A. Kozhevnikov, and M.S. Fee, 
   ``An ultra-sparse code underlies the generation of neural sequences
   in a songbird'', 
   Nature \textbf{419}, 65--70 (2002).

\bibitem[Hauschild and Jentzel (2001)]{Hauschild.Jentzel}
   T. Hauschild and M. Jentzel,
   ``Comparison of maximum likelihood estimation and chi-square
     statistics applied to counting experiments'', 
   Nucl.\ Instrum.\ Meth.\ \textbf{457}, 384--401 (2001).

\bibitem[Henze et al.\ (2000)]{Henze}
   D.A. Henze, Z. Borhegyi, J. Csicsvari, A. Mamiya, K.D. Harris, and
   G. Buzs\'aki,
   ``Intracellular Features Predicted by Extracellular Recordings in
   the Hippocampus In Vivo'',
   J. Neurophysiol.\ \textbf{84}, 390--400 (2000).

\bibitem[Ison et al.\ (2011)]{ison} 
  M.J. Ison, F. Mormann, M. Cerf, C. Koch, I. Fried, and R.Q. Quiroga,
  ``Selectivity of pyramidal cells and interneurons in the human
  medial temporal lobe'',
  J. Neurophysiol.\ \textbf{106}, 1713--1721 (2011).

\bibitem[Konkle et al.\ (2010)]{Konkle}
  T. Konkle, T.F. Brady, G.A. Alvarez, and A. Oliva,
  ``Conceptual Distinctiveness Supports Detailed Visual Long-Term Memory
  for Real-World Objects'' 
  J. Exp.\ Psychol.\: Gen.\ \textbf{139}, 558--578 (2010).

\bibitem[Kozhevnikov and Fee \ (2007)]{HVC-RAX}
   A.A. Kozhevnikov and M.S. Fee,
   ``Singing-Related Activity of Identified HVC Neurons in the Zebra Finch'',
   J. Neurophysiol.\ \ \textbf{97}, 4271--4283 (2007).

\bibitem[Lee, et al. \ (2004) ]{lee.rao}
	I. Lee, G. Rao, and J.J. Knierim,
	``A Double Dissociation between Hippocampal Subfields: Differential Time Course of CA3 and CA1 Place Cells for Processing Changed Environments'', Neuron, Vol. 42, 803--815, (2004)
	
\bibitem[Lennie (2003)]{metabolism}
  P. Lennie, ``The Cost of Cortical Computation'', Curr.\ Biol.\
  \textbf{13}, 493--497 (2003). 

\bibitem[Lyons (1986)]{Lyons}
   L. Lyons,
   ``Statistics for Nuclear and Particle Physicists''
   (Cambridge University Press, 1986).

\bibitem[Miller et al.\ (2013)]{miller.2013}
  J.F. Miller, M. Neufang, A. Solway, A. Brandt, M. Trippel, I. Mader,
  S. Hefft, M. Merkow, S.M. Polyn, J. Jacobs, M.J. Kahana,
  A. Schulze-Bonhage,
  ``Neural Activity in Human Hippocampal Formation Reveals the Spatial
  Context of Retrieved Memories'',
  Science \textbf{342}, 1111--1114 (2013).

\bibitem[Mizuseki and G. Buzs\'aki (2013)]{mizuseki.buzsaki}
  K. Mizuseki and G. Buzs\'aki,
  ``Preconfigured, Skewed Distribution of Firing Rates in the
  Hippocampus and Entorhinal Cortex'',
  Cell Rep. \textbf{4}, 1--12 (2013).

\bibitem[Mormann et al.\ (2008)]{Mormann}
  F. Mormann, S. Kornblith, R.Q. Quiroga, A. Kraskov, M. Cerf,
  I. Fried, and C. Koch,
  ``Latency and Selectivity of Single Neurons Indicate Hierarchical
  Processing in the Human Medial Temporal Lobe'', 
  J. Neurosci.\ \textbf{28}, 8865--8872 (2008).

\bibitem[Olshausen and Field (2004)]{sparse.coding.Olshausen.Field}
   B.A. Olshausen and D.J. Field,
   ``Sparse Coding of Sensory Inputs'',
   Curr.\ Opinion Neurobiol.\ \textbf{14}, 481--487 (2004).

\bibitem[Olshausen and Field (2005)]{Olshausen.Field}
   B.A. Olshausen and D.J. Field,
   ``How Close Are We to Understanding V1?'',
   Neural Comput.\ \textbf{17}, 1665--1699 (2005);
   ``What is the other 85\% of V1 doing?'',
   in ``23 problems in systems neuroscience'',
   J.L. van Hemmen and T.J. Sejnowski (eds.)
   (Oxford University Press, 2006).

\bibitem[Pedreira, Martinez, Ison, and Quian Quiroga (2012)]{spike.sorting.issue}
   C. Pedreira, J. Martinez, M.J. Ison and R. Quian Quiroga,
   ``How many neurons can we see with current spike sorting
   algorithms?'',
    J. Neurosci.\ Meth.\ 211, 58--65 (2012).

\bibitem[Plaut, McClelland \ (2010)]{Plaut_McClelland_2010a}
  D.C. Plaut and J.L. McClelland,
  ``Locating Object Knowledge in the Brain: Comment on Bower's (2009)
  Attempt to Revive the Grandmother Cell Hypothesis'',
  Psychol.\ Rev.\ \textbf{117}, 284--288 (2010).

\bibitem[Plaut, McClelland \ (2010)]{Plaut_McClelland_2010b}
  D.C. Plaut and J.L. McClelland,
  ``Postscript: Parallel Distributed Processing in Localist Models
  Without Thresholds'', 
  Psychol.\ Rev.\ \textbf{117}, 289--290 (2010).

\bibitem[Quiroga, et al. \ (2007)]{Quiroga.2007}
  R.Q. Quiroga,
  ``Decoding Visual Inputs From Multiple Neurons in the Human Temporal Lobe'',
	J.\ Neurophysiol.\ \textbf{98}, 1997--2007 (2007).

\bibitem[Quiroga (2012)]{concept.cells}
  R.Q. Quiroga,
  ``Concept cells: the building blocks of declarative memory
  functions'', 
  Nat.\ Rev.\ Neurosci.\  \textbf{13}, 587--597 (2012).

\bibitem[Quiroga, Kreiman \ (2010)]{Quiroga_Kreiman_2010a}
  R.Q. Quiroga and G. Kreiman,
  ``Measuring Sparseness in the Brain: Comment on Bowers (2009)'',
  Psychol.\ Rev.\ \textbf{117}, 291--297 (2010).

\bibitem[Quiroga, Kreiman \ (2010)]{Quiroga_Kreiman_2010b}
  R.Q. Quiroga and G. Kreiman,
  ``Postscript: About Grandmother Cells and Jennifer Aniston Neurons''
  Psychol.\ Rev.\ \textbf{117}, 297--299 (2010).

\bibitem[Quiroga et al.\ (2005)]{GMC}
  R.Q. Quiroga, L. Reddy, G. Kreiman, C. Koch, and I. Fried,
  ``Invariant visual representation by single neurons in the human
  brain'',
  Nature \textbf{435}, 1102--1107 (2005).

\bibitem[Rolls (2013)]{Rolls.2013}
  E.T. Rolls, 
  ``The mechanisms for pattern completion and pattern separation in
  the hippocampus,'' 
  Front.\ Syst.\ Neurosci.\ \textbf{7}, 74 (2013).

\bibitem[Rolls and Tovee (1995)]{Rolls.Tovee.1995}
  E.T. Rolls and M. Tovee,
  ``Sparseness of the Neuronal Representation of Stimuli in the Primate Visual Cortex'',
  J.\ Neurophys.\ \textbf{73}, 713--726 (1995).

\bibitem[Rolls and Treves (2011)]{Rolls.Treves.2011}
  E.T. Rolls and A. Treves,
  ``The neuronal coding of information in the brain'',
  Prog.\ Neurobiol.\ \textbf{95}, 448--490 (2011).

\bibitem[Shoham et al.\ (2006)]{dark.matter}
  S. Shoham, D.H. O'Connor, and R. Segev,
  ``How silent is the brain: is there a ``dark matter'' problem in
  neuroscience?'',
  J. Comp.\ Physiol.\ A \textbf{192}, 777--784 (2006)

\bibitem[Standing (1973)]{Standing}
  L. Standing, ``Learning 10,000 pictures'', Q. J. Exp.\
  Psychol.\ \textbf{25}, 207--222 (1973).

\bibitem[Thompson and Best(1989)]{silent.cells}
   L.T. Thompson and P.J. Best,
   ``Place Cells and Silent Cells in the Hippocampus of
   Freely-Behaving Rats'',
   J. Neurosci.\ \textbf{9}, 2382--2390 (1989).

\bibitem[Treves and Rolls(1991)]{Treves.Rolls91}
   A. Treves and E.T. Rolls,
   ``What determines the capacity of autoassociative memories in the
   brain?'', 
   Network \textbf{2}, 371--397 (1991).

\bibitem[Waydo et al.\ (2006)]{Waydo}
  S. Waydo, A. Kraskov, R Quian Quiroga, I. Fried, and C. Koch,
  ``Sparse Representation in the Human Medial Temporal Lobe'', 
  J. Neurosci.\ \textbf{26}, 10232--10234 (2006).

\bibitem[Wickelgren (1969)]{Wickelgren.1969}
  W.A. Wickelgren, 
  ``Learned specification of concept neurons'',
  Bull.\ Math.\ Biophy.\ \textbf{31}, 123--142 (1969). 

\bibitem[Wickelgren (1992)]{Wickelgren.1992}
  W.A. Wickelgren, 
  ``Webs, Cell Assemblies, and Chunking in Neural Nets'',
   Concepts in Neuroscience \textbf{3}, 1--53 (1992). 

\bibitem[Wickelgren (1999)]{Wickelgren.1999}
  W.A. Wickelgren, 
  ``Webs, Cell Assemblies, and Chunking in Neural Nets: Introduction'',
   Can.\ J. Exper.\ Psych.\ \textbf{53} 118--131 (1999).


\end{thebibliography}
\end{document}